\documentclass{aastex}
\usepackage{emulateapj5}
\usepackage{apjfonts}

\newcommand{\nuthin}{\nu_{\rm thin~disk}}
\newcommand{\nuthick}{\nu_{\rm thick~disk}}
\newcommand{\nustellar}{\nu_{\rm stellar~halo}}
\newcommand{\nudark}{\nu_{\rm dark~halo}}
\newcommand{\nudisk}{\nu_{\rm disk}}
\newcommand{\nuhalo}{\nu_{\rm halo}}
\newcommand{\nthin}{n_{0,{\rm thin~disk}}}
\newcommand{\nthick}{n_{0,{\rm thick~disk}}}
\newcommand{\nstellar}{n_{0,{\rm stellar~halo}}}
\newcommand{\ndark}{n_{0,{\rm dark~halo}}}

\shorttitle{WD in the Groth Strip}

\begin{document}

\title{A Proper Motion Survey for White Dwarfs with the Wide Field Planetary
Camera 2}

\author{
   C.A.~Nelson\altaffilmark{1,2},
   K.H.~Cook\altaffilmark{1,3},
   T.S.~Axelrod\altaffilmark{4},
   J.R.~Mould\altaffilmark{5},
   C.~Alcock\altaffilmark{3,6}
}

\altaffiltext{1}{Lawrence Livermore National Laboratory, Livermore, CA 94550\\
    Email: {\tt cnelson, kcook@igpp.ucllnl.org}}
\altaffiltext{2}{Department of Physics, University of California, Berkeley,
        CA 94720}
\altaffiltext{3}{Center for Particle Astrophysics, University of California,
        Berkeley, CA 94720}
\altaffiltext{4}{Research School of Astronomy and Astrophysics,
        Mount Stromlo Observatory, Cotter Road, Weston, ACT 2611, Australia\\
 Email: {\tt tsa@mso.anu.edu.au}}
\altaffiltext{5}{National Optical Astronomy Observatory, 950 N. Cherry Ave,
        Tucson, AZ, 85726\\
        Email: {\tt jrm@noao.edu}}
\altaffiltext{6}{Department of Physics and Astronomy, University of
        Pennsylvania, Philadelphia, PA, 19104-6396\\
        Email: {\tt alcock@hep.upenn.edu}}

\begin{abstract}
We have performed a search for halo white dwarfs as high proper motion objects
in a second epoch Wide Field Planetary Camera 2 image of the Groth-Westphal
strip.  The survey covers 74.8 arcmin$^2$, and is complete to $V\sim26.5$.  We
identify 24 high proper motion objects with $\mu > 0.014^{\prime\prime}$/y.
Five of these high proper motion objects are identified as strong white dwarf
candidates on the basis of their position in a reduced proper motion diagram.
We also identify two marginal candidates whose photometric errors place them
within $\sim1\sigma$ of the white dwarf region of the reduced proper motion
diagram.  We create a model of the Milky Way thin disk, thick disk and stellar
halo and find that this sample of white dwarfs is clearly an excess above the
$\leq 2$ detections expected from these known stellar populations.  The
origin of the excess signal is less clear.  Possibly, the excess cannot be
explained without invoking a fourth galactic component: a white dwarf dark
halo.  Previous work of this nature has separated white dwarf samples into
various galactic components based on kinematics; distances, and thus
velocities, are unavailable for a sample this faint.  Therefore, we present a
statistical separation of our sample into the four components and estimate the
corresponding local white dwarf densities using only the directly observable
variables, $V$, $(V-I)$, and $\vec{\mu}$. For all Galactic models explored,
our five white dwarf sample separates into about 3 disk white dwarfs and 2
halo white dwarfs.  However, the further subdivision into the thin and thick
disk and the stellar and dark halo, and the subsequent calculation of the
local densities are sensitive to the input parameters of our model for each
Galactic component.  Using the lowest mean mass model for the dark halo and
the 5 white dwarf sample we find $\nthin = 2.4^{+0.7}_{-0.6}\times10^{-2}$
pc$^{-3}$, $\nthick = 0.0^{+7.6}\times10^{-4}$ pc$^{-3}$, $\nstellar =
0.0^{+7.7}\times10^{-5}$ pc$^{-3}$, and $\ndark =
1.0^{+0.4}_{-0.4}\times10^{-3}$ pc$^{-3}$.  This implies a 7\% white dwarf
halo and six times the canonical value for the thin disk white dwarf density
(at marginal statistical significance), but possible systematic errors due to
uncertainty in the model parameters likely dominate these statistical error
bars.  The white dwarf halo can be reduced to $\sim1.5$\% of the halo dark
matter by changing the initial mass function slightly.  The local thin disk
white dwarf density in our solution can be made consistent with the canonical
value by assuming a larger thin disk scaleheight of 500 pc.  
\end{abstract}

\section{Introduction}

The  microlensing results towards the Large Magellanic Cloud
\citep{alc97, alc00, lass00} generated much interest in the possibility of white
dwarfs (WD) as significant contributors to the Galactic halo dark matter.  The most
recent results suggest a most likely MACHO fraction of 20\% and a most likely
MACHO mass between 0.15 and 0.9 $M_\odot$ \citep{alc00}. Other MACHO candidates such as
brown dwarfs, M dwarfs, and neutron stars are excluded, respectively, on the
basis of the most likely MACHO mass, direct star counts which suggest that
their contribution to the total mass is insignificant \citep{gou98} and
entirely unacceptable nucleosynthesis yields from their main sequence
precursors \citep{car94}.

White dwarfs are well-known low luminosity stars, and have been extensively
surveyed, for instance by \citet{leg98} and \citet{kno99}. A very recent
report \citet{maj01} suggests the scale height for thin disk white dwarfs
may be higher than previously thought, and that consequently the total
number of thin disk white dwarfs is also higher than previously thought.

White dwarfs as dark matter pose their own set of problems, as their main
sequence progenitor may produce more metals  (He, C, N) than observed
\citep{fie00}.  These chemical evolution constraints can possibly be avoided
by assuming a non-standard initial mass function \citep{cha96,cha99} and,
perhaps, lower metal yields from Z=0.0 zero age main sequence progenitors.
Recent results \citep{mar01} suggest that the first-dredge up does not take
place for zero metallicity stars with $M\gtrsim1.2~M_{\odot}$.
Further, the second dredge-up is suppressed for zero metallicity stars with
$M\lesssim2.1~M_{\odot}$ and only brings CNO to the surface for
$2.7~M_{\odot}\lesssim M \lesssim 8.3~M_{\odot}$.  Finally, thermal pulses on
the asymptotic giant branch which would normally bring carbon
to the surface may not occur \citep{cha99,mar01}.

Interest in WD as directly detectable dark matter intensified with the
suggestion that ancient hydrogen atmosphere WD evolve towards \textit{bluer}
optical colors as they cool \citep{han98}, remaining detectable in the $V$
band for many Gyr longer than previously assumed. This provided a plausible
explanation for some of the faint blue objects in the Hubble Deep Field (HDF),
two of which were reported to have proper motions consistent with an
interpretation as ancient halo white dwarfs \citep{iba99}.

Although the HDF moving objects were determined to be false detections
\citep{ric01}, excitement over the possibility of halo white dwarfs was
renewed with the results of \citet{opp01} who claim detection of 38 halo white
dwarfs, constituting at least 2\% of the Galactic halo dark matter.
\citet{opp01} present their results in the form of a plot of the galactic
radial (U) and rotational (V) velocities of each WD and superimpose 1 and 2
sigma contours for the expected locations of the thick disk and halo
components of the Milky Way.  White dwarfs lying outside the 2 sigma contours
of the thick disk are assumed to belong to a halo population.

\citet{rei01} provide an alternate interpretation of the \citet{opp01}
results, arguing the velocity distribution is more consistent with the
high-velocity tail of the thick disk. Furthermore, by comparing the placement
of the halo candidates along a fiducial WD evolution track in a
color-magnitude diagram, \citet{han01} notes that the \citet{opp01} halo
population seems to have an age distribution similar to the standard thin disk
population.  \citet{han01} finds that it is difficult to make the age
distribution of this sample consistent with common assumptions about thick
disk star formation (a simple burst at early times), and even more difficult
to achieve consistency with some sort of truly ancient halo population.
\citet{koo01} extend this argument, noting that the thick disk and halo WD
populations (as divided by \citet{opp01}) are indistinguishable in terms of
luminosity, color and apparent age.

\citet{koo01} undertake a more sophisticated analysis of the \citet{opp01}
sample, calculating the contribution of the thick disk and halo using a maximum
likelihood analysis.  They find a local number density of thick disk WD of
$n_{0,{\rm thick~disk}}= 1.8\pm0.5 \times10^{-3}$ pc$^{-3}$ and a local number
density of halo WD of $n_{0,{\rm halo}}= 1.1^{+2.1}_{-0.7}\times10^{-4}$
pc$^{-3}$.  The halo density is about 5 times higher then previously expected
\citep{gou98}, but constitutes only $\sim$ 0.8\% of the dark halo density, at
least an order of magnitude smaller than the MACHO density implied by the
microlensing results.  

\citet{rey01} also provide a reanalysis of the \citet{opp01} sample.  These
authors use a comprehensive description of Galactic stellar populations to
provide a simulation of the \citet{opp01} data, including the detection limits
in proper motion, magnitude and color.  They conclude that thick disk white
dwarfs with standard local densities are sufficient to explain the
\citet{opp01} sample.

The wide spread in age of the \citet{opp01} sample has led several parties to
suggest that these WD originated in the thin disk, but were subsequently accelerated
to much higher velocities.  \citet{koo01} suggest a mechanism to eject WD into
the halo with the required speeds of $\sim200$ km/s through the orbital instability
of triple systems. \citet{dav01} proposes another binary driven mechanism in which
the secondary of a tight binary system (the WD progenitor) is ejected at high velocity
when the primary explodes as a supernova.

The \citet{opp01} results are based on a very wide survey (10\% of the sky)
with a bright limiting magnitude ($R_{\rm lim} \lesssim 20$).  In this work we
take the opposite approach, examining a very small area of the sky (20 Wide
Field Planetary Camera 2 fields), but probing to a very deep limiting magnitude ($V_{\rm lim} \sim
27$).  Based on the \citet{opp01} results which are $\sim90$\% proper motion
limited \citep{koo01}, we would not expect the white dwarf luminosity 
function to rise suddenly beyond their detection limit.  However, the \citet{opp01}
survey cannot exclude the existence of an ancient ``dark halo'' white dwarf
population of age $>> 10$ Gyr.

In fact, since \citet{opp01} use the \citet{ber97} survey of young
white dwarfs to derive a linear relationship between absolute magnitude and
color, they assume that all detected white dwarfs are younger than the beginning
of the cooling turnoff towards bluer colors in the color magnitude diagram. In
their filters, this color turnoff occurs at a temperature of around 2500 K, or 
a white dwarf age of about 13-14 Gyr. Since they observe very few white dwarfs
near the color turnoff this is probably a good, if limiting, assumption.

Although we initially explore an intuitive analysis of our sample along the
same lines as \citet{opp01}, in our final analysis we avoid making any
assumptions about the age of our sample and estimate the local densities of
disk and halo white dwarfs based solely on the observed properties of our
sample: apparent magnitude, color and proper motion vector on the sky.

This work is structured as follows.  In \S2 we describe the observation
and reduction of each epoch, our procedure for matching stars between epochs,
the selection of significant high proper motion objects and the results of
our completeness tests.  In \S3 we describe our selection criteria for
white dwarfs and examine their kinematic properties in a manner similar to
\citet{opp01}.  In \S4 we model various components of the Milky Way and
compute the number of white dwarf detections we expect in our survey. Then,
using only the directly observable properties, we create a method
to statistically separate our sample into the various Milky Way components.
We conclude in \S5.

\section{Groth Strip Observations and Reductions}

First epoch images of the ``Groth-Westphal strip'' \citep{gro94} were taken by
the Wide Field Planetary Camera 2 (WFPC2) in the F606W and F814W filters
between March 7, 1994 and April 8, 1994.  This dataset consists of 27 adjacent
WFPC2 fields taken at high galactic latitude ($l \sim 96^{\circ}$, $b \sim
61^{\circ}$).  Each first-epoch field covers 4.4 arcmin$^2$, consists of 4 raw
exposures per filter and has a combined exposure time of 4400s in the standard
Hubble Space Telescope (HST) I-band (F814W) and 2800s in a narrow band V (F606W).

A re-imaging of 17 fields of the Groth strip, comprising 74.8 arcmin$^2$, was
taken by the WFPC2 in a single filter (F606W) in March, 2001, giving us a
baseline of 7 years between epochs.  Each second-epoch field consists of 6 raw
exposures and has a combined exposure time of 4200s.

\subsection{Photometry}

Because the Groth strip lies at such high galactic latitude, the fields are
nearly devoid of bright stellar objects, containing $\sim 4$ stars of 
$V<26$ per WFPC2 chip.  This makes it difficult to compute
exact transformations between the multiple raw images of the same field,
especially as the raw images are heavily contaminated by incident cosmic ray
particles.  Thus, in combining the raw images we assume that the requested
pointing offsets between raw images (commonly called dithers) were accurate
and the images were combined using these offsets.  Generally, WFPC2 pointings
are accurate to $\sim 0.1$ WF pixels, so this assumption does not add serious
error to the procedure.  The first epoch images are all taken at the same
pointing (un-dithered) so the raw images are combined with zero offset.  The
second epoch images are dithered by non-integer values.  We list the
fields and times observed in Table 1.

After the raw images were aligned, all subsequent processing and reductions
were performed using the HSTphot package made publicly available by Andrew
Dolphin \citep{dol00}.  Each raw image is first masked using the bad pixel
mask.  The aligned raw images are then combined and cleaned of cosmic rays.
For each pixel, the cosmic ray cleaning algorithm computes the median pixel
value, a photon counting uncertainty ($\sigma$) based on the median value,
the exposure time, read noise and gain.  The uncertainty is then given an
additional contribution added in quadrature to account for the registration
uncertainty.  Pixel values which were more then $3\sigma$ from the median
value were rejected and the remaining pixel values were averaged.

Point-spread fitting (PSF) photometry was performed on the combined images
using the HSTphot package. The HSTphot package includes a library of PSFs
constructed with the Tiny Tim software.  For the F814W filter it also includes
default residuals to the ``perfect'' Tiny Tim PSFs.  In addition, HSTphot also
allows the PSF to be adjusted based on measurements made on suitable stars in
the user's program frames.  In the Groth strip frames, there are very few stars
with enough counts to act as suitable PSF stars. Attempts to adjust the PSF
using available stars resulted in unusual looking PSFs, although the 
final results were insensitive to the PSF used (see below). Thus, in the final
reductions we forced HSTphot not to adjust its PSF based on our program
frames.  For the F814W band reductions we allowed it to use the default residuals.
According to the HSTphot documentation, not adjusting the PSF based on the
program stars can create a systematic error of
up to $\pm$0.1 magnitudes for faint stars.  We allowed HSTphot to internally
compute aperture corrections.  The HSTphot magnitudes were converted to
Landolt $V$ and $I$ using the synthetic transformations of Table 10 from
\citet{hol95b}.

HSTphot also includes an object classification routine which divides objects
into stars, possible unresolved binaries, single-pixel cosmic rays or extended
objects.  HSTphot was designed for crowded stellar fields and the use of this
software on fields of mostly galaxies is certainly an application beyond its
intended purpose.  Thus, it is unsurprising that the object classification
routine was often unreliable, classifying objects as stellar which, upon
inspection seemed more likely to be galactic.  Regardless of the output
classification, HSTphot fits a stellar PSF to all objects and the output
magnitude and centroids are based on this fit.  The quality of its output is
thus directly related to whether or not the object was actually a point 
source.

\begin{figure}
\epsscale{0.5}
\plotone{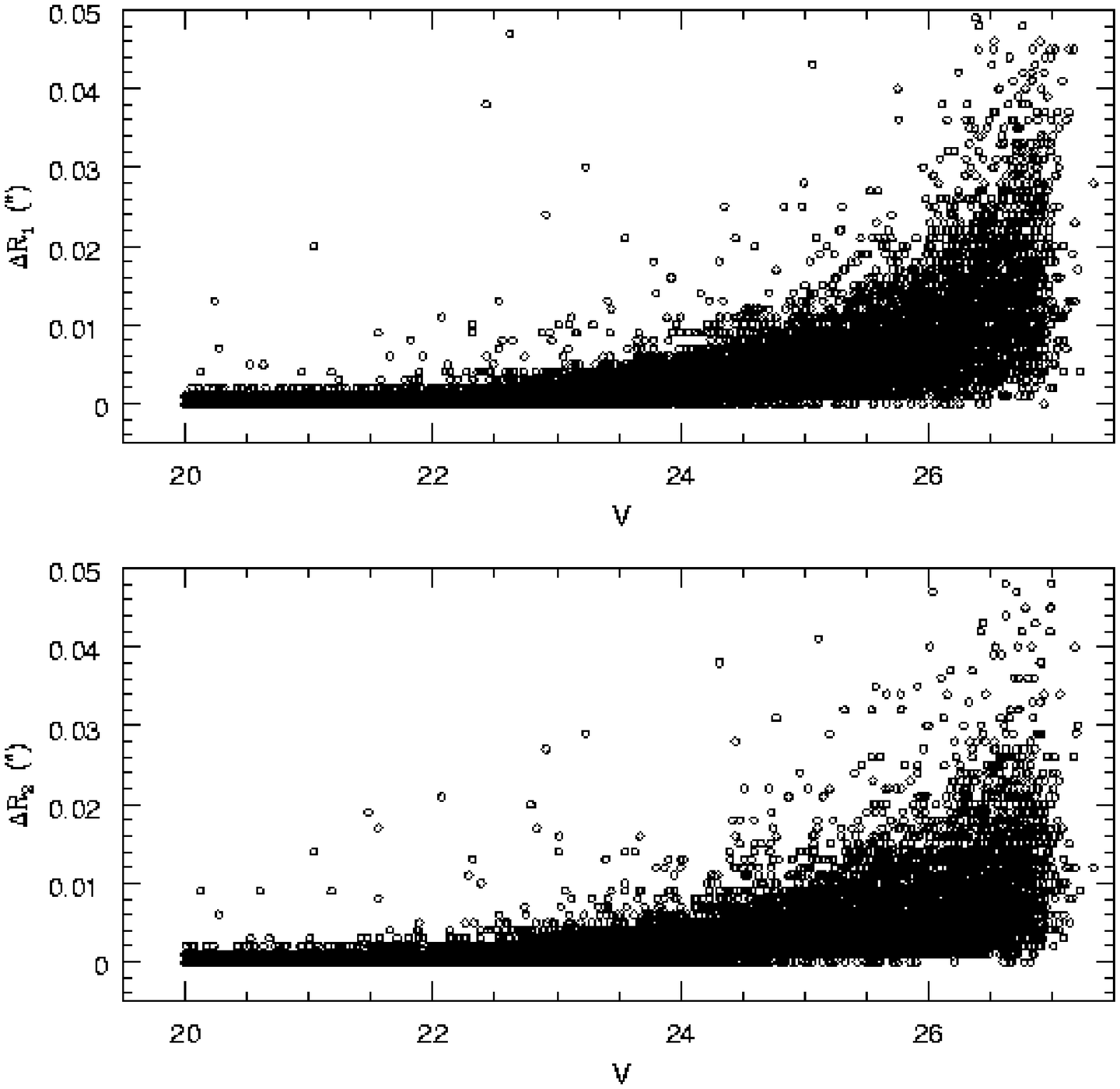}
\caption{The uncertainty in the centroids computed by HSTphot as a function of 
$V$ magnitude. The uncertainty is estimated using our artificial star tests.
$\Delta R_1$ indicates the difference between the input and recovered centroid
for artifical stars in the epoch 1 frames.  $\Delta R_2$ indicates the same
quantity for epoch 2.}
\label{centroid}
\end{figure}

We estimate a total photometric error for faint objects ($V\lesssim24$)
of $\sim0.15$ mag with contributions from both the non ideal PSF and photon
counting error.  We can use the artificial star tests (described in
\S~\ref{cmp}) to estimate the centroid accuracy of HSTphot for stellar objects
by computing the difference between the input and recovered centroid.
The centroid uncertainty as a function of magnitude for all artifical stars is
shown in Figure~\ref{centroid}.  This uncertainty is underestimated in the
sense that it tests the ability of HSTphot to centroid objects with a perfect
PSF.  The artificial star tests add stars to frames with the same PSF (plus
random noise) with which they are recovered.  Since this PSF is not perfect
for the real stars in our frame, the centroid uncertainty for real stars is
likely substantially higher.  The greater uncertainty in the centroids for
real stars is reflected in final uncertainty in the coordinate
transformation between epochs (see \S~\ref{match}) which is much larger than
the HSTphot centroid uncertainty of Figure~\ref{centroid}.

\subsection{Matching stars between epochs}
\label{match}

Accurate computation of the transformation between two epochs of a field
is best performed using many, bright, point-like sources.  
Bright stellar sources are in short supply
in the Groth strip and we are forced to compromise to some extent.

The first step in computing transformations between two WFPC2 frames is to
correct for the geometric distortion of the chips.  There are several sets of 
distortion coefficients available including one specifically optimized
for the first epoch Groth strip \citep{rat97}, but the most widely employed are from
\citet{hol95a}.  Each set of coefficients can also put the four WFPC2 chips
on a single ``global'' coordinate frame.  Since the accuracy of our
transformations is limited by the low number of sources per field,
a global coordinate system would seem to be a useful tool.  We experimented
with transformations based on several different global coordinate systems
and always found the residuals of these transformations to depend heavily
on the X or Y position of the star on the chip.  
This problem arises not from the distortion coefficients themselves but 
from the fact that the geometric transformation of WFPC2 has a small time
dependence, primarily in the interchip separation which can vary by as much
as 150 mas. Thus, there exists no global distortion solutions
that work well on both the first epoch (1994) and second epoch (2001)
data.  This limited us to computing local transformations on a chip by
chip basis using the \citet{hol95a} distortion coefficients.  

Since the reduction of this data (in April 2001) a new set of distortion
coefficents has become available (WFPC2 Instrument Science Report 2001-10).
We have tested a reduction routine in which we use the \cite{hol95a} 
distortion coefficients for our epoch 1 data and the new distortion
coefficients for our epoch 2 data.  We find no significant improvement in our
results. The dominant source of uncertainty in our transformations arises
from our imperfect PSF and the small number of objects in our  fields, not the
distortion correction used.

Because we were unable to use global transformations, we were also unable
to use the data in the Planetary Camera (PC).  The extremely small area
of the PC ($\sim0.3$ square arcminutes) rarely encompassed
enough objects in the frame with which to compute local transformations.
For consistency, we simply ignore all PC data.

Before performing the transformations, we developed an object classification
routine based on the SExtractor software \citep{ber96}.  
SExtractor identifies objects above a certain threshold and
then utilizes a neural network classification routine to assign each object a
decimal number between 0 and 1 where 0 indicates a stellar (point-source)
object and 1 indicates a galactic (extended-source) object.  Initially, we
computed transformations based only on objects that were classified as stellar by
SExtractor with the appropriate centroids taken from the HSTphot output.  For
many chips there simply were not enough such stellar objects to compute
transformations.  Thus, we developed a routine where transformations were
computed using stellar objects and ``small, round'' galaxies.  After
much experimentation we found that better transformations were obtained from
HSTphot centroids (PSF fitting centroids) for both stars and galaxies than
from using the SExtractor centroids for the galaxies.  Computing the 
coordinate transformations is an iterative procedure, in which we
remove objects with residuals greater then 3 sigma before computing the
final transformations. Final transformations
computed are a full 6 parameter fit allowing for translation, rotation and
change of scale.  
Our transformations are computed assuming that the vector sum of the proper
motions over all stars and galaxies was zero. 

\begin{figure}
\epsscale{0.5}
\plotone{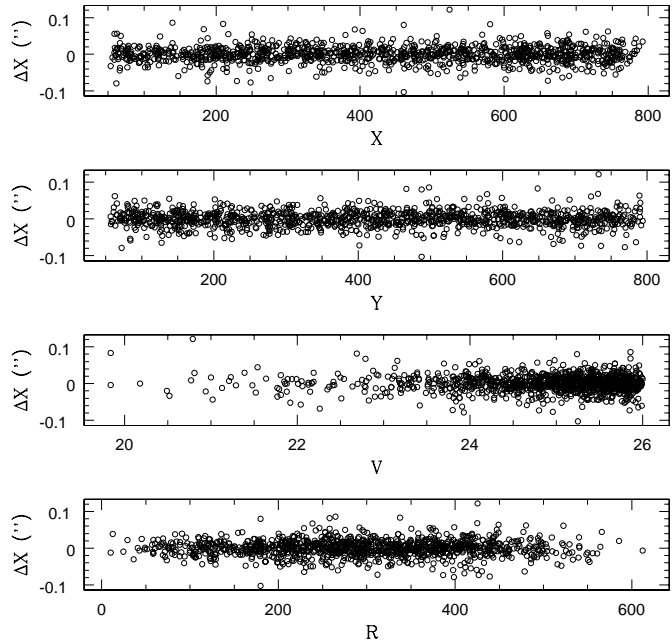}
\caption{Residuals in the X coordinate, $\Delta X$, to the transformations
between epochs.  The top two panels show $\Delta X$ plotted versus the X and
Y position on the chip.  The third panel shows the residuals as a function of
$V$ magnitude.  The bottom panel shows the residuals as a function of radius
from the distortion center of the chip.  Residals are plotted in units of
arcseconds.}
\label{res_x}
\end{figure}

\begin{figure}
\epsscale{0.5}
\plotone{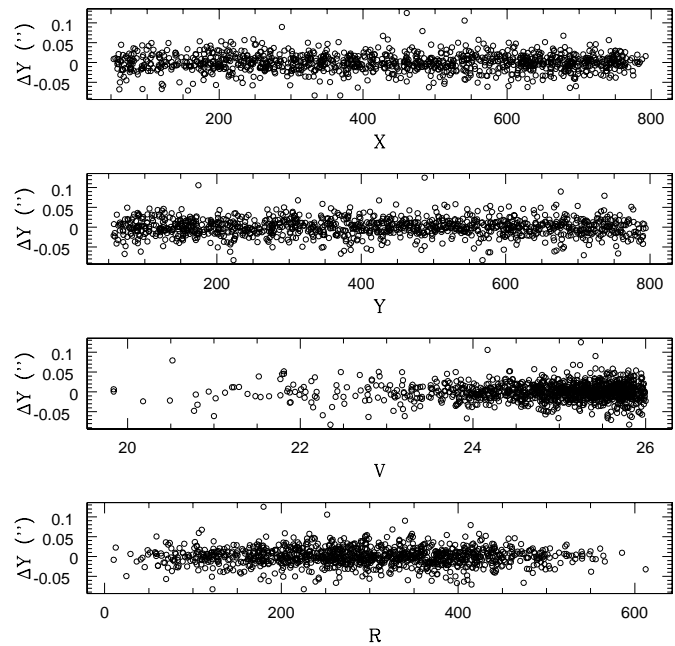}
\caption{As in Figure~\ref{res_x} but for residuals in the Y coordinate,
$\Delta Y$.}
\label{res_y}
\end{figure}

In Figures~\ref{res_x} and~\ref{res_y}, we plot the residuals to the
coordinate transformation between epochs for all Groth fields, $\Delta X$
and $\Delta Y$ as a function of X and Y position on the chip, magnitude,
and radius from the distortion center at $(X,Y)=(354,344)$.
The residuals show no obvious dependence on position
indicating that we have corrected for the geometric distortion of the
chips.  The residuals show some dependence on magnitude, increasing towards
fainter magnitudes.  We compute the transformations and residuals only for 
stars with $V < 26$ although we allow fainter stars to appear in the final
list of matched stars for a frame.  The main reason for this is that below $V
\sim 26.5$ real stars in epoch 2 start to get matched with noise in epoch 1
(epoch 2 is $\sim0.5$ magnitude deeper).  Since the final high proper motion
star lists are inspected by hand (see below), it is acceptable to include
ill-matched stars in the final matched lists, but it is crucial that we
minimize their occurrence in the list of stars used to compute the
transformations.

\begin{figure}
\epsscale{0.5}
\plotone{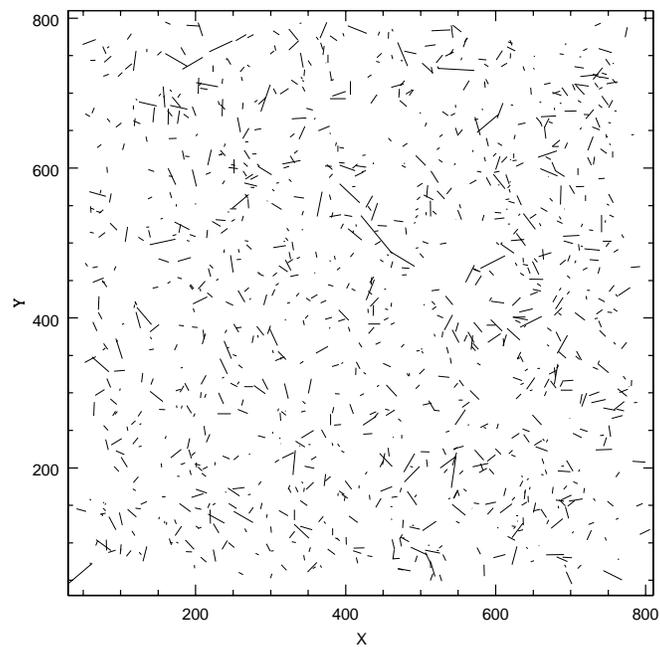}
\caption{Vector plot of the residuals to the transformations as a function
of position on the chip. The residuals are multiplied by a factor of 40 to
improve visibility.}
\label{vec_res}
\end{figure}

The transformation residuals are used to compute standard deviations in x and
y, $\sigma_x$ and $\sigma_y$, and then we define the total standard deviation
as $\sigma = \sqrt{\sigma_x^2+\sigma_y^2}$.  In Figure~\ref{vec_res} we show a
vector plot of the residuals for all stars used to compute the transformations in
all fields as a function of position on the WFPC2 chip.
The sigma for all fields combined is $\sigma = 0.031^{\prime
\prime}$. 

Our final uncertainty in the coordinate transformation, $\sigma =
0.031^{\prime\prime}$ is a substantially worse result then the limits reached by
Anderson \& King (2000).  Our analysis is limited in comparison to their work
in several respects.  First, we cannot derive an effective PSF as we have
virtually no suitable PSF stars and the first epoch was undithered.  Without
dithering and a large sample of PSF stars one cannot hope to adequately sample
the pixel phase space and derive an effective PSF.  Second, we have very few
point sources per frame (a median value of 4 per chip) with which to derive
transformations.  Although we supplement the point source centroids with
centroids of small galaxies, our ability to derive accurate transformations
suffers as a result of the non point source PSF of the galaxies.  We note that
our estimate of 4 stars per chip is in line with the previous first epoch
Groth strip reductions of \citet{gou97} who find an average
of 3.1 M dwarfs per chip.

\subsection{High Proper Motion Objects}

Once the transformations were computed, lists of matching objects with $V \leq
27$ were made. Objects were matched to their
nearest neighbor to a maximum of $0.75^{\prime \prime}$ and were considered
matched only if the $V$ magnitude in each epoch matched to within 0.5 mag.  
An object was considered to have a significant movement between epochs 
and is hereafter known as a high proper motion object if
the residual to the transformation for that object was above the maximum
value of 3$\sigma$ for that field or $0.1''$ (1 pixel).  That is, the proper motion
limit is defined as 
\begin{equation}
\mu_{\rm lim} \cdot \Delta t = {\rm min}(3\sigma,0.1^{\prime \prime})
\end{equation}
In order to minimize confusion while matching stars between epochs, we chose
an upper proper motion limit of 7.5 pixels, or $\mu_{max} \sim
0.1^{\prime\prime}$/yr.  Each high proper motion object was inspected by hand
to verify that it was a plausible match between epochs and possessed a stellar
PSF.

\begin{figure}
\epsscale{0.5}
\plotone{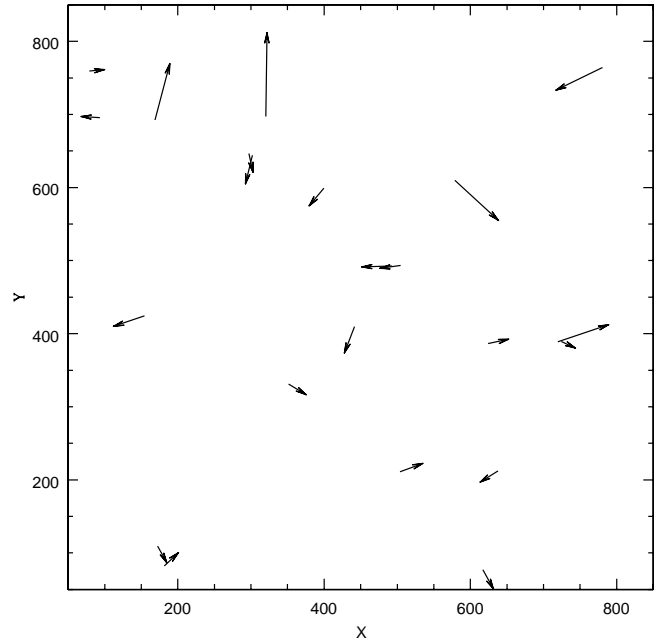}
\caption{Vector proper motions for the high proper motion objects as a function
of position on the WFPC2 chip. The proper motions are multiplied by
a factor of 20 to improve visibility.}
\label{xy}
\end{figure}

In Figure~\ref{xy}, we plot the vector proper motion as a function of the
location on the WFPC2 chip for all high proper motion (HPM) objects.  Averaged
over all chips and all fields, the expected distribution of high proper motion
objects over the face of the chip should be roughly uniform.  We do not see
any obvious departures from non-uniformity.  Most importantly, we do not see
any clustering of barely significant HPM objects towards the edges of the chip
where the geometric distortions are most severe.

We repeated the full analysis using the adjustable PSF option in HSTphot.
While a few barely significant movers in the constant PSF reduction were
not significant in the adjustable PSF option and vice versa, the list of 
HPM objects was similar.  All final WD candidates discussed below
appeared in both the adjustable and constant PSF reductions.

\subsection{Completeness tests}
\label{cmp}

We performed completeness tests by adding artificial stars to the $V$ band
images from both epochs.  We compute the completeness fraction as a function
of both magnitude and proper motion. First, coordinates for the star in epoch
2 were chosen randomly.  We then compute the corresponding position in the
epoch 1 frame using the derived transformations and apply a random offset
(proper motion) of $-3 \leq \Delta x \leq 3$, $-3 \leq \Delta y \leq 3$ 
pixels before adding the star to the epoch 1 frame.  If the transformed position does
not lie on the epoch 1 frame, the star is not added to either epoch frame.
Because of this, the effective area of our survey is only the area in which we
have complete overlap between the two epochs.  We create 25 frames each
containing 20 artificial stars in each camera. 

\begin{figure}
\epsscale{0.5}
\plotone{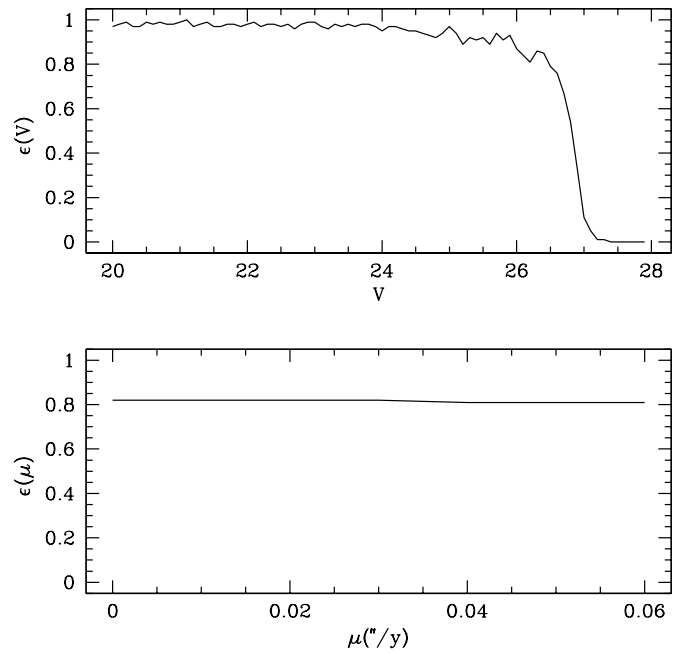}
\caption{The fraction of objects detected as a function of V magnitude,
$\epsilon(V)$ and proper motion, $\epsilon(\mu)$.  Results are averaged over
all Groth strip fields.}
\label{eff}
\end{figure}

The artificial frames are then run through the standard photometric analysis.
We do not recompute the transformations between epochs for the
artificial frames.  In this sense, the proper motion axis of the completeness
test is artificial - since we add stars with the correct transformation, of
course we recover them with the same transformation.  Our completeness
fraction is thus uniform in proper motion and really depends only on the
$V$ magnitude. The completeness fraction as a function of $V$ magnitude and
proper motion is shown in Figure~\ref{eff}.  Since we compute our completeness
fraction only in $V$ (and not $I$) the calibrated magnitudes of the artificial
stars are computed without a color term.  

We compute our completeness as a function of $\mu$ only to 4.25 pixels, yet
in our analysis we accept objects in our samples with proper motions of up
to 7.5 pixels.  We chose the 4.25 pixels to be large enough to include the
range of all observed high proper motion objects, yet small enough to allow
good coverage in $\mu$ in a reasonable amount of computation time.  As 
discussed above, our efficiencies do not depend on $\mu$ and we expect no
dependence to arise if our efficiency calculation were extended to 7.5 pixels.

We expect that the detection efficiency should fall to exactly zero at 
$V=27.0$ as we do not allow stars with $V<27.0$ to appear in the final
list of matched stars.  The detection efficiency is not quite zero in
this plot because we calculate the efficiency as a function of 
$V_{input}$ and we recover a very small fraction of stars that have
$V_{input} > 27.0$ but are recovered at $V < 27.0$ and thus appear in 
the final matched lists. 

Our completeness fraction falls rapidly to zero at $V\sim26.5$, suggesting
a limiting magnitude $V_{\rm lim}\sim 26.5$.

\section{White dwarf candidates}

\subsection{Selection criteria and reduced proper motions}
\label{selection}

After an initial period of nearly blackbody cooling towards longer
wavelengths, ancient white dwarfs experience a color turnoff and start to cool
towards bluer wavelengths \citep{han99}. An example cooling curve of a 0.6 $M_\odot$
WD is shown in Figure~\ref{track}.  Thus, we expect to detect halo white
dwarfs as blue objects with high proper motions.  One way to isolate
such candidates is on the basis of a reduced proper motion versus color diagram.

\begin{figure}
\epsscale{0.5}
\plotone{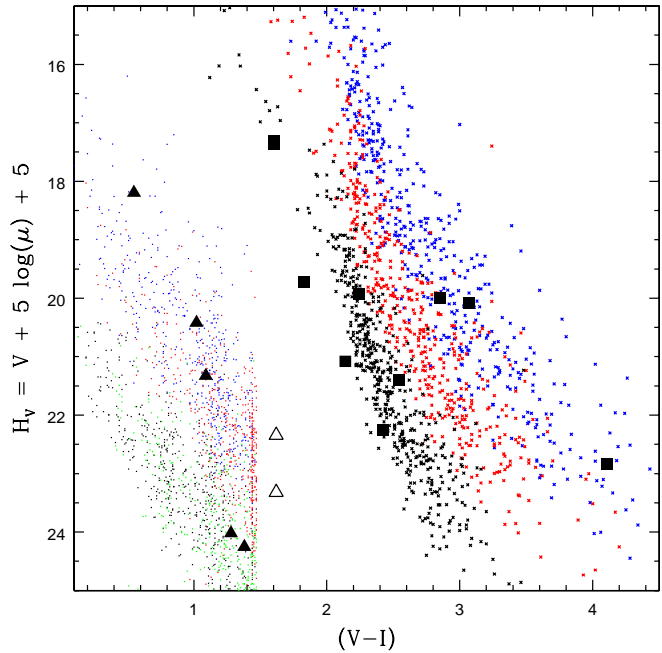}
\caption{Reduced proper motion diagram for all HPM objects in the Groth Strip.
The strong WD candidates are shown as filled black triangles, the nearby
HPM main sequence stars are shown as filled black squares.  The open triangles
indicate the two marginal white dwarf candidates discussed in \S\ref{selection}. 
The small dots indicate a simulated sample of detectable WD in the thin disk
(blue), thick disk (red), stellar halo (black) and dark halo (green).  The
small crosses indicated a simulated sample of low mass main sequence stars
($M<0.9M_\odot$) from the thin disk (blue), thick disk (red) and stellar halo
(black).}
\label{rpm}
\end{figure}

In Figure~\ref{rpm} we show the reduced proper motion diagram for all the 
significantly moving (high
proper motion) objects in the Groth strip.  The reduced proper motion is
defined as $H_V = V +
5\log(\frac{\mu}{^{\prime\prime}/{\rm yr}}) + 5 = M_V +
5\log(\frac{v_t}{\rm{km/s}}) - 3.378$.  For an ensemble of objects with
similar transverse velocities, $v_t$, an $H_V$ versus color diagram is an estimate
of the color-absolute magnitude diagram where the magnitude axis
has been shifted by some arbitrary constant.  In Figure~(\ref{rpm}) we
show the Groth strip HPM objects as large triangles and squares.

On the left hand side of Figure~\ref{rpm} we show a random sample of detectable
white dwarfs from the thin disk (blue dots), thick disk (red dots), stellar
halo (green dots) and dark halo (black dots).  The parameters determining the
geometry and stellar content of each 
Galactic components and our method for determining detectability will be
discussed further in \S~\ref{simulations}.  We note that the five HPM objects
marked as filled black triangles fall over plausibly dense regions of the
simulated white dwarf dots.

On the right hand side of Figure~\ref{rpm} we show a random sample of
detectable low mass main sequence stars $(M<0.9M_\odot)$ from the thin disk
(blue crosses), thick disk (red crosses) and stellar halo (black crosses).
The parameters of the Galactic components are the same as for the WDs and
we use the low mass main sequence tracks from \citet{bar98}.
The HPM objects shown as filled black squares fall on plausibly dense regions
of the simulated low mass main sequence star crosses.  If we had included higher mass
main sequence stars, the density of points would fall off much slower at
small $H_V$.

On the basis of this reduced proper motion diagram we identify the filled
black triangles as strong white dwarf candidates and the filled black
triangles as main sequence stars.  The $V$ magnitude, $(V-I)$ color
and proper motion, $\mu$, of the white dwarf candidates are shown in Table 2.

The two stars shown as open triangles have  $V>26$ and $\mu = 1.1 \mu_{\rm
lim}$ and $\mu=1.3 \mu_{\rm lim}$, respectively, and fall in an unpopulated
region of the reduced proper motion diagram.  Since we calculate the
3$\sigma$ significant proper motion limit $\mu_{\rm lim}$ with stars of $V<26$
and the proper motions of these faint stars are only barely above the proper
motion limit, it may be the case that these two stars are not HPM objects at
all, but simply spurious detections at very faint magnitude.  However, it may
also be the case that these two stars are moving objects, and are in fact
white dwarfs near the color-turnoff.  To move these two objects 
into the white dwarf region of the reduced proper motion diagram would
require an error of $0.15$ magnitudes in $(V-I)$ in either
our photometry or the theoretical white dwarf cooling tracks. Since an error
of this magnitude is plausible, we will later consider the possibility that
these two objects may belong to our white dwarf sample.  We include these two
objects in Table 2 as marginal candidates.

As noted by \citet{maj01}, quasars (QSOs) can be an important contaminant of the
white dwarf region of the reduced proper motion diagram.  \citet{maj01}
investigate this issue in detail and conclude that the soundest way to
to eliminate QSOs from a white dwarf sample is to choose only objects which
move with some certain minimum proper motion. They suggest that QSOs are
almost certainly detected at proper motion of less than four times the
uncertainty (eg. $\mu < 4\sigma$) and that objects with proper motions above
this limit are more certainly WD.  The condition $\mu > 4\sigma$
is satisfied for all of our candidates (see Table 2).

Hereafter we shall refer to WD candidates 1-5 from Table 2 as our 5 WD
sample.  Since we have the most confidence in this sample we will devote
most discussion to it.  We will also discuss the possibility
that the marginal candidates do belong to our WD sample and will refer
to WD candidates 1-7 from Table 2 as our 7 WD sample.  For the sake of 
completeness, we will also discuss the possibility that all of our 
faint candidates are in error and that only WD 1, 3 and 4 from Table 2 
belong to our WD sample.  This will be referred to as the 3 WD sample.

We emphasize that these are not spectrally confirmed WD.  With the exception
of the bright candidate at $V=20.7$ these objects are too faint to allow for
spectral confirmation with present technology. We also note that the proper
motion of these candidates has not been confirmed with a third epoch. A previous
WFPC2 detection of high proper motion WD by \citet{iba99} has recently been
withdrawn after a third epoch revealed the reported proper motion to be a 
spurious measurement \citep{ric01}.  
We note that all our candidates have
substantially higher counts in both epochs than the second epoch image of
\citet{iba99}.

\subsection{Kinematics}

\begin{figure}
\epsscale{0.5}
\plotone{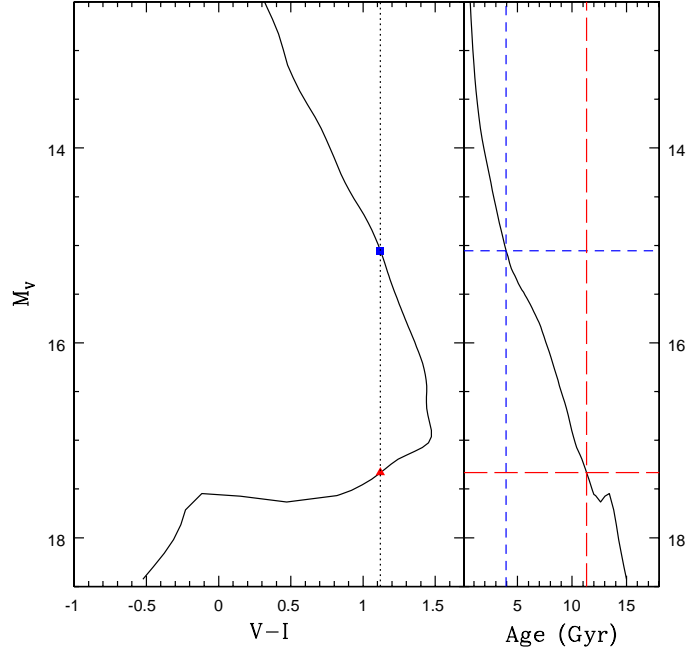}
\caption{Cooling curves for a 0.6 $M_\odot$ WD from the models of
\citep{ric00}.  As  WD cool $H_{2}$ molecules form in the atmosphere and
create an opacity which is very strong in the near infra-red and forces the flux
to emerge at bluer wavelengths. Thus, in a $V$ versus $(V-I)$ color magnitude
diagram a WD initially cools along a blackbody curve, but eventually the WD
start to become blue again.  Because of this bend in the cooling track, a
given $(V-I)$ color intersects the cooling curve at two points (black dotted
line) , giving both a ``young'' (blue square and short-dashed lines) and an
``old'' solution (red triangle and long-dashed lines.}
\label{track}
\end{figure}

In order to determine the transverse velocity of each candidate, we must first
determine its distance, or, equivalently, its absolute magnitude.  In
principle, this may be done by finding the intersection of the color of each
candidate with an absolute magnitude versus color track, such as that shown in
Figure~\ref{track}.  However, because of the color turnoff, the $(V-I)$ color
does not uniquely determine $M_V$.  Instead, for each candidate we have a
``young'' and an ``old'' solution as illustrated in
Figure~\ref{track}. For each solution we may calculate the distance, $d =
10^{\frac{V-M_V+5}{5}}$ and the transverse velocity, $v_t = \mu d$, where $\mu$
is the proper motion.  Both the young and old solutions for the WD cooling
time, distance, and transverse velocity are shown in Table 2.  We include
error bars on the derived quantities (cooling time, distance, transverse
velocity and absolute magnitude) by calculating the maximum and minimum
values obtained while I adjusting the $V$ magnitude and $(V-I)$ color by
one unit of error (0.15 mag for $V$ and 0.20 mag for $(V-I)$).
We note that the cooling time is not the total age of the WD as it does not
include the main sequence lifetime of the WD precursors.

In a simplistic sense, the marginal candidates have no solution for
the derived parameters since they do not intersect a white dwarf track
in $(V-I)$ color.  We assume that it is plausible that the observed 
$(V-I)$ color is wrong by some small amount and that they may just barely
intersect the reddest point of the theoretical white dwarf track.
Since we assume they only barely intersect the tip of the color turnoff,
they cross the white dwarf track at only one point and have only one solution.

Also, we note that even if all the ``old'' solutions are correct, most of our
WD are located at least one thin disk scaleheight (300 pc) above the plane.
We also include the proper motion in the direction of increasing Galactic
longitude, $\dot{l}$, and latitude, $\dot{b}$ in Table 2.  The spherical
coordinate system employed implies that $\mu = \sqrt{ (\dot{l}\cos{b})^2 +
(\dot{b})^2 }$.

This two solution ambiguity is not present in brighter, ground based samples
where the distance can be determined with a photometric distance relation.
Such a relation is essentially a linear fit to the young branch of a WD
track in a color-absolute magnitude diagram.  One such relation is derived in
\citet{opp01} using the \citet{ber97} sample of nearby white dwarfs for
which parallax distances are available.  The \citet{opp01} sample employs this
relation without much fear of ambiguity because in their filters ($B_J$, $R59F$)
the color turnoff does not occur until a cooling time of $\sim$13.5 Gyr.  In contrast,
in our filters, the color turnoff occurs at the relatively short cooling time of
$\sim$10 Gyr.

In \citet{opp01}, the white dwarfs are roughly divided into thick disk and
halo components by examining the location of the white dwarfs on a $(U,V)$
diagram, the velocity components towards the Galactic center and in the
direction of Galactic rotation, respectively.  The authors are able to
calculate space velocities without information about the radial velocites by
assuming that the velocity out of the plane, $W$, is zero (the mean value for
all stars in the disk).  Since the \citet{opp01} sample is near the South
Galactic cap, this is not a bad assumption in the mean.

We encounter several difficulties in determining the space
velocities of our sample: the two solution distance ambiguity, the lack of
radial velocities and a lower Galactic latitude ($b\sim60^{\circ}$) at which
it is more difficult to make assumptions about the contribution of the radial
velocity to
the space velocities.  In the next section we attempt a division of our sample
into various Galactic components without making assumptions about the
distances and velocties of our WD candidates.

\section{Simulations}
\label{simulations}

In this section, we interpret our results in a method independent of 
the white dwarf distance and 
assumptions about the young or old solutions. Instead, we examine
our sample using only the directly observable variables 
$(V,V-I,\vec{\mu)}$. We begin in \S4.1 by determining the number of 
white dwarf detections we expect from known stellar populations and
compare this to our observed number.  In \S4.2 we consider a possible contribution
from a white dwarf dark halo.  \S4.1 and \S4.2 define our reference model of
the Galaxy. In \S4.3 we attempt a
statistical division of our sample into several galactic components. In \S4.4
we discuss alternative models of the Galaxy. In \S4.6 we discuss the possiblity of alternative
WD samples.

\subsection{Expected number of white dwarfs from known stellar populations}
\label{known}

First, we construct a simulation of the local WD populations of the MW,
calculate the expected number of WD detections from each population, and
compare this number with the Groth strip observations. In principle, given
a model for each Milky Way component, we are able to predict
the number of expected WD detections because we have an accurate estimate of
our detection efficiency as a function of both proper motion and magnitude.

In the simulation, we explore three known stellar components of the Milky Way:
the thin disk, thick disk and stellar halo.  Estimating the number of WD we
can detect from each component requires the knowledge of many model
parameters.  These parameters include $n_0$, the local number density of WD,
$\rho(R,z)$ or $\rho(r)$, the shape of the density function (in
cylindrical $(R,z)$ or spherical $(r)$ coordinates), $v_a$, the
asymmetric drift of the population, and ($\sigma_U$,$\sigma_V$,$\sigma_W$),
the velocity dispersions towards the Galactic center, in the direction of
Galactic rotation and out of the plane towards the Galactic North Pole.  We
also need a WD luminosity function suitable for the age distribution and 
initial mass function (IMF) of each component.  For all components, we assume that the
number density of WD has the same shape as the mass density,
i.e.~$n(R,z) = \rho(R,z) / \bar{m}_{WD}$ where $\bar{m}_{WD}$ is the mean WD
mass. 

We model the Milky Way thin and thick disks as double exponentials of form
\begin{equation}
n_{\rm disk} (R,z) = 
n_{c}~\exp\left(  - \frac{R}{R_d} - \left| \frac{z}{z_d} \right| \right)~,
\end{equation}
where $R$ and $z$ are cylindrical coordinates with origin at the Galactic
center, $R_d$ is the scale length, $z_d$ is the scale height and $n_c$ is the
central density.  The central density is determined from the local density by
$n_c = n_{0,{\rm disk}} \exp{(R_0/R_d)}$ and $R_0=8.5$ kpc is the radius of the
solar circle.  

For the thin disk, we use the parameters $R_d = 4$ kpc and $z_d = 0.3$ kpc.
(Note that \citet{maj01} argue for a larger scale height for the thin disk.)
The thin disk velocity dispersions and asymmetric drift are taken to be
$(\sigma_U,\sigma_V,\sigma_W)=(34, 21,18)$ km/s and $v_a = 6$ km/s
\citep{bin98}.  For the thick disk we use the parameters $R_d = 4$ kpc and
$z_d = 1.0$ kpc. The thick disk velocity dispersions and asymmetric drift are
taken to be $(\sigma_U,\sigma_V,\sigma_W)=(61, 58,39)$ km/s and $v_a = 36$
km/s \citep{bin98}.  

The Milky Way stellar halo density is given by 
\begin{equation}
n(r)_{\rm stellar~halo} = \nstellar (r/R_0)^{-3.5}~,
\end{equation}
where $r$ is Galactocentric radius and $R_0 = 8.5$ kpc is the Galactocentric
radius of the Sun \citep{giu94}.  The velocity dispersions
are taken from \citet{chi00} to be $(\sigma_U,\sigma_V,\sigma_W)=(141,106,94)$
km/s and we assume an asymmetric drift of $v_a = 220$ km/s.

Our initial assumptions for $n_0$ are guided by previous observations. The
local thin disk WD density has been determined in two independent samples to
be $n_{0,{\rm thin}} \approx 4\times10^{-3}$ pc$^{-3}$ \citep{leg98, kno99}.
We scale the thick disk to the thin disk such that $n_{0,{\rm thin}}/
n_{0,{\rm thick}}=42$ as in \citet{alc00}. The stellar halo WD density is
estimated from subdwarf star counts and a standard initial mass function in
\citet{gou98} as $n_{0,{\rm stellar~halo}}=2.2\times10^{-5}$ pc$^{-3}$ (for
$\bar{m}_{WD}=0.6$ $M_\odot$).

Our choice of a suitable luminosity function for each component is  more
arbitrary. We create luminosity functions using the white dwarf cooling curves
of \citet{ric00}, main sequence lifetimes from \citet{gir00} and initial main
sequence mass to final white dwarf mass relations from \citet{van97}.  For the
known stellar populations, we assume a Saltpeter initial mass function with
$\alpha = -2.35$ and we make the approximation that when stars leave the main
sequence they instantly become white dwarfs.  The cooling curves of
\citet{ric00} are appropriate to white dwarfs with a mixture of carbon and
oxygen cores and a hydrogen surface layer.

\begin{figure}
\epsscale{0.5}
\plotone{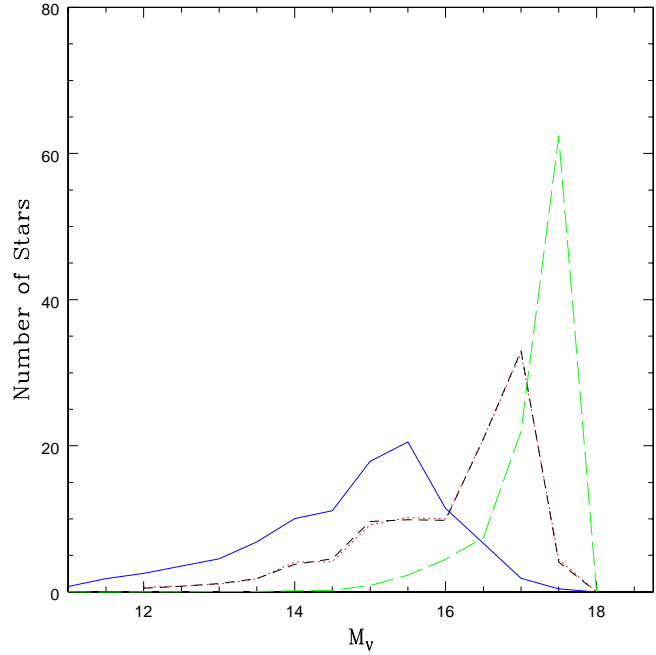}
\caption{White dwarf luminosity functions for the thin disk (blue solid line),
thick disk (red dotted line), stellar halo (black short-dashed line)
and dark halo (green long-dashed line).
The dark halo luminosity function is created using the 96IMF1 initial
mass function parameters.}
\label{lf}
\end{figure}

For the thin disk we assume a metallicity $Z = 0.02$ and a uniform star
formation rate over the past 11 Gyr. \citet{gil00} asserts that all thick disk
and field halo stars formed within 1--2 Gyr of the onset of star formation,
and that there is no detectable age interval between the formation of the field
halo stars and the formation of the thick disk.  Thus, we model both the thick
disk and stellar halo as populations with a uniform star formation rate between
11--12 Gyr with initial metallicity $Z=0.004$. The luminosity functions we 
generate are shown in Figure~(\ref{lf}).

The geometry and stellar content of the thin disk, thick disk and stellar halo
described above will be referred to hereafter as our reference model for the
known stellar populations.

We begin the simulation by generating a large sample of WD from each
component with parameters drawn from models described above.  For each
simulated WD, we draw a distance along the line of sight,
$d<d_{\rm max}$, weighted appropriately according to the density distribution.
We choose $d_{\rm max} = 5.0$ kpc, a distance large compared to our maximum WD
detection limit.  We randomly choose space velocities,
$(U,V,W)$ from a Gaussian distribution with the appropriate velocity
dispersions.  We also draw a luminosity, $M_V$, according to the component
luminosity function.  A color, $(V-I)$, is then assigned according to the
tracks for a 0.6 $M_\odot$ WD from \citet{ric00}.

Next, we transform variables into the observable quantities.  The space velocities
are transformed into a radial velocity, $v_r$, and velocities in the direction
of Galactic longitude and latitude, $(v_l,v_b)$, by inverting the following
system of equations:
\begin{equation}
\label{U}
U = U_\odot + v_r\cos{b}\cos{l} -  v_b\sin{b}\cos{l} - v_l\sin{l}~,
\end{equation}
\begin{equation}
\label{V}
V = V_\odot + v_r\cos{b}\sin{l} -  v_b\sin{b}\sin{l} + v_l\cos{l}~,  
\end{equation}
\begin{equation}
\label{W}
W = W_\odot + v_r\sin{b} +v_b\cos{b}~.
\end{equation}
In these equations,
$v_l=d~\mu_l\cos{b}$, $v_b=d~\mu_b$, and $(U_\odot,V_\odot,W_\odot)$ =
(10, 5.2, 7.2) km/s is the motion of the Sun with respect to the LSR
\citep{deh98}. The
transverse velocity on the sky is then $v_t = \sqrt{v_b^2 + v_l^2}$ and the
proper motion is given by
\begin{equation}
\left( \frac{\mu}{^{\prime\prime}/{\rm y}} \right) = 0.211
\left( \frac{v_t}{{\rm km/s}} \right)
\left( \frac{d}{{\rm pc}} \right)^{-1}~.
\end{equation}
The apparent magnitude is determined by $V = M_V + 5\log(d/{\rm pc}) - 5$.
Finally, we draw a random number, $r$ between 0.0 and 1.0 and declare the WD
to be detectable if $r < \epsilon(V)$ and $\mu_{\rm lim} \sim
0.014^{\prime\prime}/{\rm y} < \mu < \mu_{\rm max} \sim
0.1^{\prime\prime}/{\rm y}$. The function $\epsilon(V)$ indicates our 
detection efficiency as a function of $V$ magnitude as plotted in
Figure~\ref{eff}.

In this way, we create large lists of detectable and non-detectable WD for
each component, $k$, from which we may determine the fraction of WD detected
in each component, $f_k$.  We then integrate  the number density along our
line of sight to determine the number of  WD, $K_k$, out to $d_{\rm max} =
5.0$ kpc.
The number of WD from this component we expect to see in the Groth strip is
then $\nu_k = f_k K_k$.

As discussed above, the number of expected WD in the Groth strip determined by
these simulations scales linearly with our assumptions about the local density
of each component. The expected number of WD detections in the Groth strip
from known stellar populations scale as 
\begin{equation}
\label{thinscale}
\nuthin = 0.5 
\left( \frac{\nthin} {4.0\times10^{-3} {\rm pc}^{-3}} \right)~,
\end{equation}
\begin{equation}
\label{thickscale}
\nuthick = 0.1
\left( \frac{\nthick} {9.5\times10^{-5} {\rm pc}^{-3}} \right)~,
\end{equation}
\begin{equation}
\label{stellarscale}
\nustellar = 0.2
\left( \frac{\nstellar} {2.2\times10^{-5} {\rm pc}^{-3}} \right).
\end{equation}
The expected number of white dwarfs from known stellar populations is 
relatively insensitive to our somewhat arbitrary choice of absolute
magnitude luminosity functions.  The insensitivity is due to the
fact that the apparent magnitude luminosity functions of each
component are actually determined more by the kinematics of each component
then the input absolute magnitude luminosity function.  For instance, as the thin
disk is the most severely proper motion limited, thin disk WDs are detected at
a mean distance much closer to the observer and thus at a much brighter
apparent magnitude.  Even a drastic adjustment to the input absolute
magnitude luminosity function such as giving the thick disk an age of 0--11 Gyr
only increases the expected contribution from the thick disk by $\sim40$\%.
Alternative disk luminosity functions are discussed in further detail in
\S~\ref{alternate_disk}.

We expect to detect a total of 0.9 white dwarfs from known stellar
populations, if our assumptions about the local white dwarf densities, $n_0$,
are accurate.  Even if we presume a maximal thick disk with
a local density scaling to the thin disk of $\sim5$ (instead of 42),
the expected total number of detections
is still less then 2. (We note that using the thin disk scale height of
\citet{maj01} increases the expected number of thin disk detections to 1.9.
This is discussed further in \S~\ref{alternate_disk}.)

Clearly, our observed sample of 5 white dwarfs is
a significant deviation from an expected number of $\sim$1--2.  However, the
source of the excess of white dwarfs is an open question.  Perhaps the
simplest explanation is that our estimate of the local white dwarf
densities $n_0$ are too small for one or more of the known stellar components.
Another possibility is that the excess of white dwarfs arises from a fourth
component: a dark halo made of ancient white dwarfs.  Below, we discuss what
we would expect to detect from such a white dwarf component. 

\subsection{Dark halo WD}
\label{dark}

Using the same procedure as for the known stellar populations, we estimate the
number of WD detections we expect if some fraction, $f_{WD}$, of the Milky Way
dark matter halo is made up of white dwarfs.  This discussion is motivated by
recent microlensing results which suggest that 8\%--50\% of the dark halo is
made up of MACHOs with a most likely MACHO mass between 0.15 and 0.9 $M_\odot$
\citep{alc00}.

The Milky Way dark halo density is modeled as
\begin{equation}
n(r)_{\rm dark} = \ndark \frac {R_0^2 + a^2}{r^2 + a^2}~,
\end{equation}
where $r$ is Galactocentric radius, $R_0 = 8.5$ kpc is the Galactocentric
radius of the Sun, and $a=5$ kpc is the halo core radius \citep{alc00}.  We
use the same velocity dispersions and asymmetric drift as for the stellar halo.  

If a substantial portion of the Milky Way dark halo is composed of WD, the
initial mass function of this population must be rather sharply peaked in
order to avoid chemical evolution constraints.   The most plausible initial
mass function is peaked at around 2 $M_\odot$ \citep{fie00}.  As an example,
we construct a dark halo luminosity function with a uniform star formation
rate  between 13.0 and 14.0 Gyr with the IMF1 of
\citet{cha96} (hereafter referred to as 96IMF1).  This IMF is of the form
\begin{equation}
\phi(m) \propto \exp^{-(\bar{m}/m)^\beta} m^{-\alpha},
\end{equation}
where $\bar{m} = 2.0$, $\beta=2.2$, and $\alpha = 5.15$ and peaks
around $\sim 1.3 M_\odot$. We estimate main
sequence lifetimes from $Z = 0.001$ stars from \citet{gir00} and assume stars
instantly become WD on leaving the main sequence. This choice of dark halo
luminosity function is also shown in Figure~\ref{lf}.  The geometry, age
distribution and 96IMF1 initial mass function described above will hereafter be
referred to as our reference model for the dark halo.

The number of expected detections from the dark halo scales as
\begin{equation}
\label{darkscale}
\nudark = 3.0 
\left( \frac
{\ndark} 
{1.6\times10^{-3} {\rm pc}^{-3}}
\right)
\end{equation}
where $\ndark = \frac{\rho_{0,{\rm dark~halo}}}{\bar{M}_{WD}} \cdot f_{WD}$, and $f_{WD}$
indicates the fraction of the total halo dark matter made up of white dwarfs.
We assume a local dark halo mass density $\rho_{0,{\rm dark~halo}} =
8.0\times10^{-3} {\rm M_\odot~ pc}^{-3}$, a mean white dwarf mass of 0.5
$M_\odot$ and scale our answer to a white dwarf halo fraction of 10\%
($f_{WD}=0.10$). 

The number of expected dark halo white dwarfs is relatively insensitive to the
age of the dark halo in the range 12-14 Gyr.  However, $\nudark$ depends
critically on the choice of IMF.  The IMF described above has the lowest mean
initial mass of all the Chabrier IMFs, implying that stars remain on the main
sequence longer, and have spent less time cooling as white dwarfs thus giving
us the brightest possible dark halo luminosity function. Using
IMF2 from \citet{cha96} (hereafter 96IMF2) which has a slightly higher mean
main sequence mass results in a slightly fainter white dwarf luminosity
function and decreases the expected number of dark halo white dwarf detections
to $\nudark\sim1$ for a 10\% WD halo ($f_{WD} = 0.1$).

\subsection{Dividing the sample: a maximum likelihood approach}
\label{division}

Taken at face value, if we accept the local white dwarf densities from known
stellar populations derived in \S~\ref{known} then the total number of WD
candidates, $N_{WD}=5$ implies $\nudark = 5 - 0.5 - 0.1 - 0.3 = 4.1$, which by
Equation~(\ref{darkscale}) implies a dark halo WD fraction $f_{WD} = 0.14$,
in line with the microlensing results of \citet{alc00}.  This
interpretation is misleading because our knowledge of the local white
dwarf densities and various other model parameters is imperfect.  It is not
clear whether the excess white dwarfs in our sample arise from 
an excess of white dwarfs from known stellar populations or whether we must
invoke the presence of a fourth dark component to explain our results.

In order to address this question, we must use more information about our
sample than simply the total number of observed white dwarfs.  Ideally, we
would approach the separation of our sample into various components with a
purely kinematic approach as in \citet{koo01}, removing the most uncertain
aspect of our models (age and initial mass functions of each component) from
the calculation all together.  However, the white dwarfs in our sample are,
with one exception, too faint to allow for spectroscopic velocity
measurements.  They are also potentially too old to allow for a photometric
distance calculation as in \citet{opp01} and too distant for parallax
measurements.  Conversely, samples which are bright enough to allow for
distance determinations are not able to exclude the presence of a truly
ancient WD dark halo (see e.g. \S 10 of \citet{koo01}).  The existence of an
ancient WD halo must be confirmed or denied with a sample in which space
velocities cannot be measured, only proper motions.

Proper motions alone will not currently allow any reasonable division of our
sample into four separate components.
In order to accomplish this we must use all the available information
for each white dwarf : $(V,V-I,\vec{\mu})$. Using the component models
described above, we may form four dimensional distribution functions in these
variables and compare the shape of these distributions with our observed
sample. The comparison takes the form of a maximum likelihood analysis in
which we determine what linear combination of distributions from each
component best fits our observed sample.  By introducing the luminosity and
color information into the analysis instead of focusing solely on kinematics,
our results will depend somewhat on our assumptions about the luminosity
functions (especially for division between the stellar and dark halo).
Since improved future knowledge about the age and initial mass
functions of each component and a larger sample size may reduce this
sensitivity, we proceed with a brief investigation of one possible method for
separation of a white dwarf sample into various components using only the
photometric and astrometric information available for faint objects.

\begin{figure}
\epsscale{0.5}
\plotone{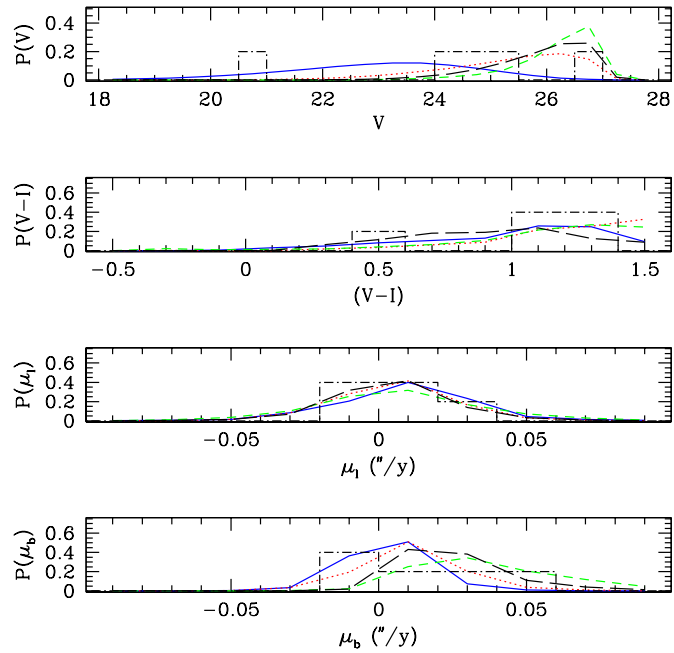}
\caption{Distributions of detectable WD in the simulations as a function of
the observable parameters $(V,V-I,\vec{\mu})$ for various Milky Way
components.  We show thin disk WD as a blue solid line, thick disk WD as a red
dotted line, stellar halo WD as black short-dashed line and the 96IMF1 dark halo WD
as a green long-dashed line.  We also overplot the distribution of observed WD
from the 5 WD sample as a black dot-dashed histogram.
}
\label{wd_distrib}
\end{figure}

In Figure~\ref{wd_distrib}, we show the one dimensional projections of the
distribution functions $P(V,V-I,\vec{\mu})$ for detectable WD in our
simulations from the thin disk (blue), thick disk (red), stellar halo (black)
and dark halo (green).  The parameters for each Galactic component are those
of our reference model described in \S~\ref{known} and \S~\ref{dark}. We also show the
distribution of the observed 5 WD  sample as a dashed black histogram.  All curves have
been normalized such that the total area under the curve is unity. The proper
motion vector has been divided into two components, $\mu_{l}$ and $\mu_{b}$,
the proper motion in the direction of increasing galactic latitude and
longitude, respectively.

The curves in Figure~\ref{wd_distrib} are the one dimensional projection of a
four dimensional cube, $P_k(V,V-I,\vec{\mu})$, in which each element of the
cube, $i$, holds the probability that a WD from each component $k$ will be
detected within some small volume element centered about $(V,V-I,\vec{\mu})$.
The cube is normalized such that the sum over all cells is unity, 
\begin{equation}
\sum_{i} P_k(V_i,V_i-I_i,\vec{\mu}_i) = 1.0~.
\end{equation}
If the number of WD drawn from a given component is $\nu_{k}$, then the
occupation number of cell $i$ is 
\begin{equation}
\lambda_{ik} = \nu_{k}~P_k(V_i,V_i-I_i,\vec{\mu_i})~.
\end{equation}
If we sum the four dimensional cubes over all components, $k$,  multiplied by
the appropriate coefficient, $\nu_k$, we create the distribution function
of white dwarfs for the entire galaxy.  The occupation number of a cell in
this cube is thus
\begin{equation}
\lambda_i = \sum_{k} \nu_{k}   P_k(V_i,V_i-I_i,\vec{\mu_i})~.
\end{equation}

We assume that the cell size is small enough that $\lambda_i<<1.0$ and that
in a given sample of observed WD, a cell will contain at most one white dwarf.
A sample of observed WD then defines two sets of cells: the set F whose cells are
filled with the observed WD, and the set E whose cells are empty.  Given a defined
set F of observed WD, the likelihood of a configuration with a given set 
of coefficients $\{\nu_k\}$ is 
\begin{equation}
L = L(\nu_k) = \left[ \sum_{i\in F} \lambda_i \right]
\left[ \sum_{i \in E} (1.0 - \lambda_{i}) \right] ~.
\end{equation}
Maximizing over the four dimensional space, we find that the maximum likelihood
occurs at ($\nuthin $, $\nuthick$, $\nustellar$, $\nudark$)$=(3.0,0.0,0.0,2.0)$
$\equiv(w_0,x_0,y_0,z_0)$.  

\begin{figure}
\epsscale{0.5}
\plotone{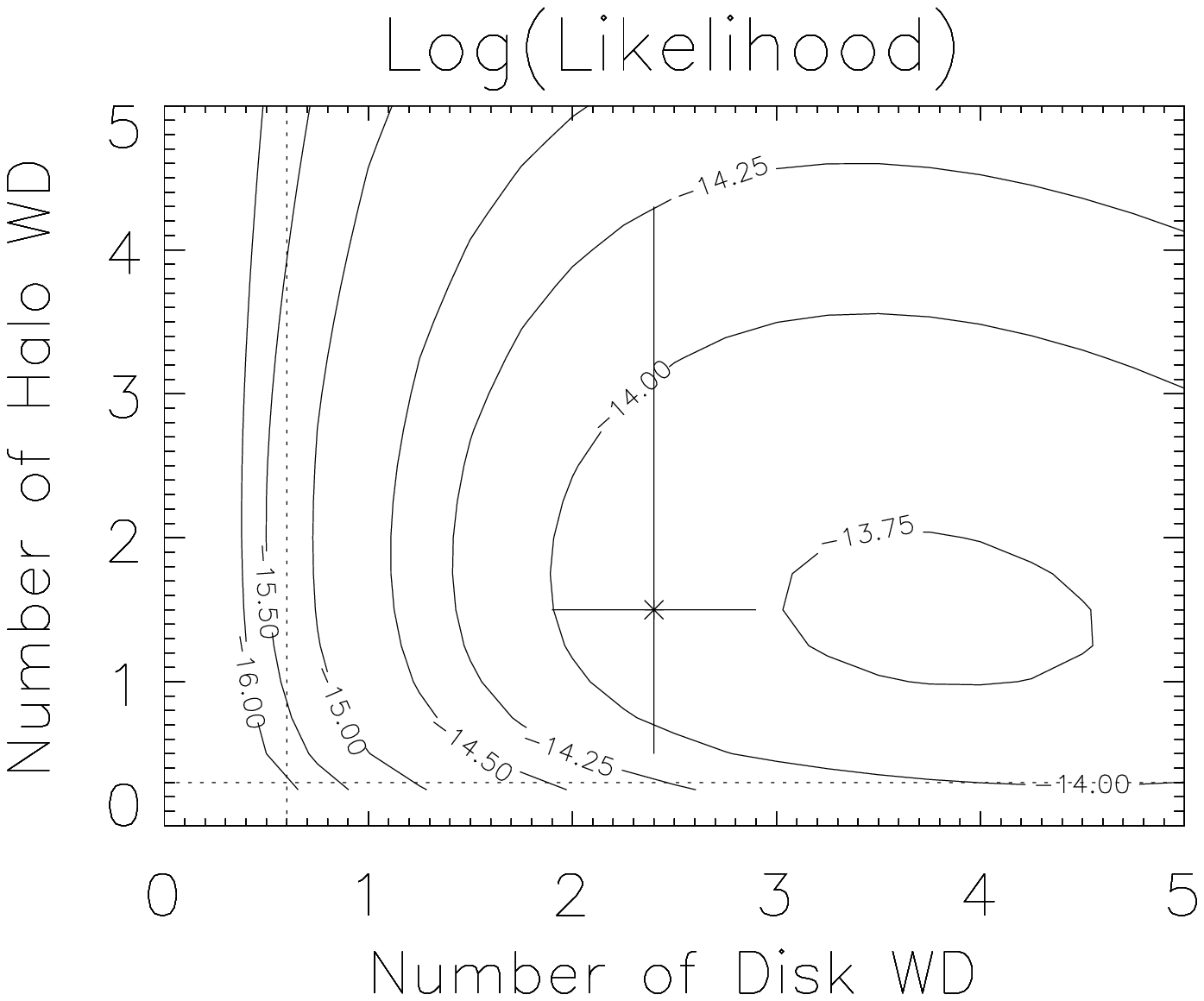}
\caption{ Likelihood contours collapsed into two dimensions.  For each point
we average all points in the four dimensional likelihood surface where
$(\nuthin + \nuthick = \nudisk)$ and $(\nustellar + \nudark = \nuhalo)$.  The
vertical dashed line shows the number of expected disk white dwarfs with the
canonical number density of a \citet{leg98} thin disk plus a thick disk scaled
by $n_{0,{\rm thin}}/n_{0,{\rm thick}}=42$ as in \citet{alc00}.  The
horizontal dashed line shows the number of expected stellar halo white dwarfs
with the canonical number density of \citet{gou98}.  Our most likely point
falls at $\nudisk=3.75$, $\nuhalo=1.5$, several orders of magnitude in
likelihood from the intersection of the two canonical values.  We also mark
with an X the number of expected white dwarfs assuming the \citet{koo01} 
number densities with an additional contribution from a \citet{leg98} thin
disk and assuming that the \citet{koo01} white dwarfs with halo kinematics
belong to the stellar halo.
}
\label{two_component}
\end{figure}

In Figure~(\ref{two_component}), we show a two dimensional projection of
the likelihood space where we sum over the thin and thick disk to form a
``disk'' component, and sum over the stellar and dark halo to form a ``halo''
component. That is, for each point in this two dimensional surface we average
all points in the four dimensional likelihood surface where $(\nuthin +
\nuthick = \nudisk)$ and $(\nustellar + \nudark = \nuhalo)$.  The vertical
dashed line shows the number of expected disk white dwarfs assuming the canonical
number densities of Equations~\ref{thinscale} and~\ref{thickscale}. 
The horizontal dashed line shows the
number of expected stellar halo white dwarfs with the canonical number density
of Equation~\ref{stellarscale}.  Our most likely point in this projection falls at
$\nudisk=3.75$, $\nuhalo=1.5$, several orders of magnitude in likelihood from
the intersection of the two canonical values.  This plot emphasizes that we
see too many WD in all components, disk and halo.

We calculate statistical error bars on the set of $\nu_k$ in the four
dimensional maximum likelihood analysis using Bayes' Theorem:
\begin{equation}
P(w,x,y,z) = P( w,x,y,z | \{w_0,x_0,y_0,z_0\}_{\rm actual}) 
\end{equation}
\begin{equation}
~~~~~~~~~~~~~~ = \frac { P(w_0,x_0,y_0,z_0|w,x,y,z)}{ \sum_{w,x,y,z} P(w_0,x_0,y_0,z_0|w,x,y,z)}
\end{equation}
The term on the left hand side is the probability of finding the maximum
likelihood value at $(\nuthin,\nuthick,\nustellar,\nudark) = (w,x,y,z)$ if the
actual values are ($\nuthin$, $\nuthick$, $\nustellar$, $\nudark$) $=
(w_0,x_0,y_0,z_00)$.  The terms on the right hand side are computed in a monte
carlo fashion.  To compute $P(w_0,x_0,y_0,z_0|w,x,y,z)$ we assume that
($\nuthin$, $\nuthick$, $\nustellar$, $\nudark$) $= (w,x,y,z)$ are the correct parameters
and form 2000 simulated data sets drawn from the co-added four dimensional
distribution function with these coefficients.  For each simulated data set we
compute the maximum likelihood point.  $P(w_0,x_0,y_0,z_0|w,x,y,z)$ is then
the fraction of the simulated data sets with input coefficients $(w,x,y,z)$
whose maximum likelihood point falls within some small volume of
$(w_0,x_0,y_0,z_0)$.

\begin{figure}
\epsscale{0.5}
\plotone{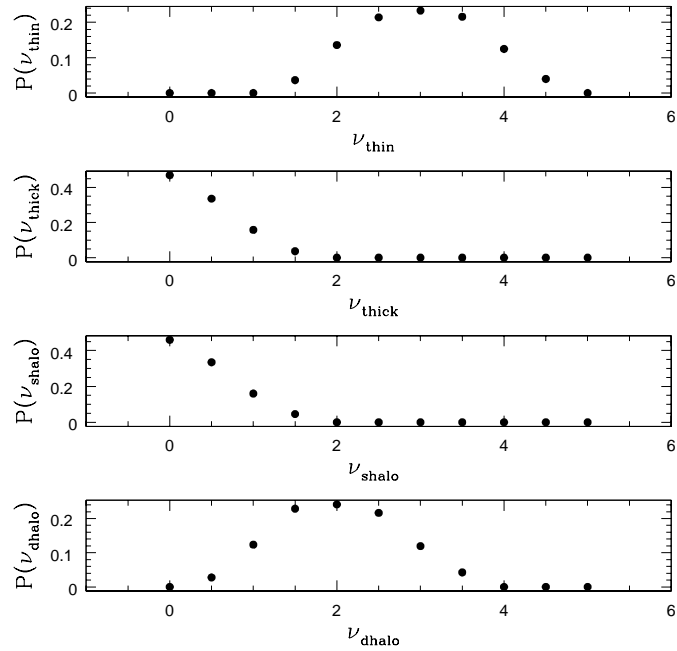}
\caption{The probability distribution functions for each parameter,
($\nuthin$, $\nuthick$, $\nustellar$, $\nudark$) determined in our maximum likelihood 
analysis for the 5 WD sample and the 96IMF1 dark halo IMF.}
\label{pdf}
\end{figure}

We have now created a four dimensional probability distribution function
$P( w,x,y,z )$ in parameter space.  We reduce this
to more intuitive one dimensional functions by summing over the remaining
variables, eg.
\begin{equation}
\label{erreq}
P( w) = \frac
{ \sum_{x,y,z} P(w_0,x_0,y_0,z_0|w,x,y,z)}
{ \sum_{w,x,y,z} P(w_0,x_0,y_0,z_0|w,x,y,z)}~.
\end{equation}
The resulting probability distribution function for $\nuthin,\nuthick,
\nustellar$ and $\nudark$ are shown in Figure~\ref{pdf}.  The curves have
been arbitrarily renormalized such that the total area under each curve is
1.0. We compute $1\sigma$ error bars as the smallest distance which encloses
69\% of the area under the curve, giving $\nuthin = 3.0^{+0.9}_{-0.8}$,
$\nuthick = 0.0^{+0.8}$, $\nustellar=0.0^{+0.8}$ and
$\nudark=2.0^{+0.8}_{-0.8}$.  The error bar calculation is extremely
computationally intensive and we estimate that the sums expressed in
Equation~\ref{erreq} are converged to $\sim$5\%. We devoted the available
CPU cycles in such a manner that the peaks of the probability distribution
functions in Figure~\ref{pdf} were calculated to greater accuracy then the
wings.  Further computation would serve to raise the wings slightly and
thus increase the error bars slightly.

These values may now be used to work backwards to observed values for the local WD
number densities using the scaling relations in
Equations~(\ref{thinscale}--\ref{stellarscale}) and~(\ref{darkscale}).   With
the error bars calculated as described above, our final analysis yields $\nthin
= 2.4^{+0.7}_{-0.6}\times10^{-2}$ pc$^{-3}$, $\nthick =
0.0^{+7.6}\times10^{-4}$ pc$^{-3}$, $\nstellar =
0.0^{+7.7}\times10^{-5}$ pc$^{-3}$, $\ndark = 1.0^{+0.4}_{-0.4}\times10^{-3}$
pc$^{-3}$.  

Given this model of the dark halo using the 96IMF1 dark halo luminosity
function, our results imply the presence of a $\sim7$\% white dwarf dark halo
(with substantial statistical error bars) and an excess of thin disk white
dwarfs. The thin disk contribution is $\sim6$ times the canonical value.
However, the statistical error bars on the thin disk contribution are quite
large - even with the error bars projected into one dimension, we are still
less then 2.5$\sigma$ from the canonical value.  It is also possible that this
spuriously high value arises because our model of the thin disk is wrong.  We
discuss the possibility of a thin disk with a much larger scale height
\citep{maj01} in \S~\ref{alternate_disk}.

\subsection{Alternative Galactic models}

We define in \S~\ref{known} and \S~\ref{dark}  one possible model for our Galaxy.
However, all parameters
in this model are uncertain to some extent.  The kinematics are
uncertain for stars as distant as our sample and the age and initial mass
functions chosen above are certainly disputable.  In this section we explore
alternative and equally credible models of the Galaxy.  Although the
modifications we explore below do change our results somewhat, our solution
always returns roughly $\sim 3 $ disk WD and $\sim 2$ halo WD.  Minor modifications
to our models and procedure shift the disk WD around between the thin and thick
disk and the halo WD between the stellar and dark halo.

In the following examples, we do not recompute statistical error bars as this
exercise is prohibitively computationally expensive.  We include this section
to emphasize that our statistical error bars of $\sim30$\% are actually small
compared to the systematic uncertainties in our models.

\subsubsection{Alternative disk models}
\label{alternate_disk}

Our models for the thin and thick disk include information about both the
geometry (density profiles, scale heights, velocity dispersions) and stellar
content (age distribution, initial mass function) of each component.  We find
that the division of our sample between the thin and thick disks is
insensitive to the age distributions of the disk components and depends
mainly on the kinematics.  

\begin{figure}
\epsscale{0.5}
\plotone{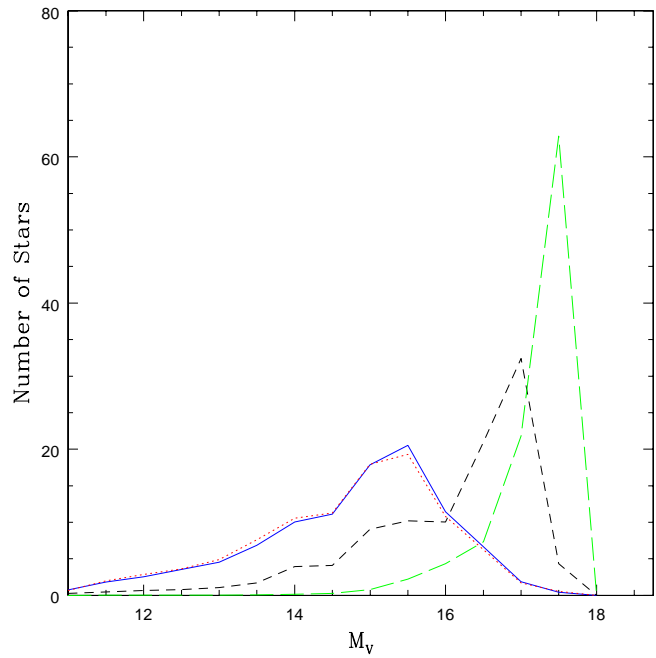}
\caption{ White dwarf absolute magnitude luminosity functions for each
Galactic component as in Figure~\ref{lf} except with a young thick disk of 
age 0--11 Gyr.}
\label{thick_young_lf}
\end{figure}

In \S~\ref{division} we describe the results of our maximum likelihood 
analysis using our reference model which has a
thin disk of age 0--11 Gyr and a thick disk of age 11--12 Gyr, modelling the
thick disk as an old population formed in a single burst.
At the opposite extreme, we may consider a thick disk whose age distribution
is identical to that of the thin disk, uniform star formation from 0--11 Gyr
ago.  This gives a thick disk absolute luminosity function nearly identical to that of
the thin disk, shown in Figure~\ref{thick_young_lf}.  (The small variation
between the thin and thick disk luminosity function is due to our different
metallicity assumptions, Z=0.02 for the thin disk and Z=0.004 for the thick
disk.)  

\begin{figure}
\epsscale{0.5}
\plotone{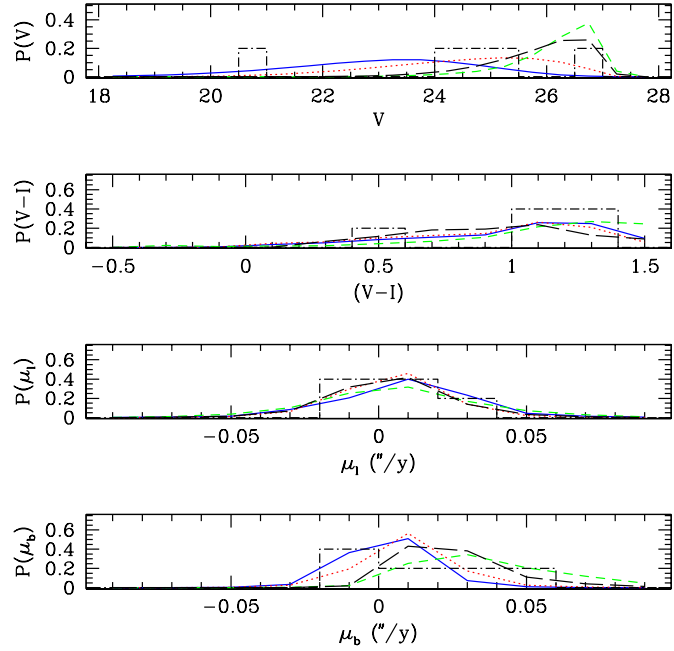}
\caption{Distributions of detectable WD in the simulations as a function of
the observable parameters $(V,V-I,\vec{\mu})$ for various Milky Way
components as in Figure~\ref{wd_distrib} but with the young thick disk
luminosity function of Figure~\ref{thick_young_lf}.}
\label{thick_young_distrib}
\end{figure}

However, despite the near identical nature of the thin and thick disk absolute
magnitude luminosity functions in Figure~\ref{thick_young_lf}, the resulting
apparent magnitude luminosity functions (shown in
Figure~\ref{thick_young_distrib}) are quite different.  Conversely, despite
the great difference in absolute magnitude luminosity function between
the 0--11 Gyr thick disk in Figure~\ref{thick_young_lf} and the 11--12 Gyr
thick disk in Figure~\ref{lf}, the apparent magnitude luminosity functions
of the thick disks in Figures~\ref{wd_distrib} 
and~\ref{thick_young_distrib} are very similar.

The apparent magnitude luminosity functions of the 0--11 Gyr and 11--12 Gyr
thick disks remain similar because the apparent magnitude 
luminosity functions are more sensitive to the kinematics than to the
input absolute magnitude functions. The thin disk has smaller velocity
dispersions and is, therefore, more severely proper motion limited than
the thick. Thus, the mean distance at which we detect a
thin disk WD is much smaller than the mean distance at which we detect a thick
disk WD. From this it follows that the detectable thin disk WD are also
brighter in apparent magnitude.

The kinematics dominate the difference in apparent magnitude luminosity
functions of the thin and thick disks to such an extent that a by-eye
comparison of the thick disk distributions in Figures~\ref{wd_distrib}
and~\ref{thick_young_distrib} shows that the difference in observable
quantities between a thick disk of age 11--12 Gyr and a thick disk of age 0--11
Gyr is slight.  Performing our maximum likelihood analysis with the 0--11 Gyr
thick disk and the other components as in our reference model,
we find, $\nthin=2.4\times10^{-2}$ pc$^{-3}$,$\nthick = 0.0$
pc$^{-3}$, $\nstellar = 0.0$ pc$^{-3}$, $\ndark = 1.0\times10^{-3}$ pc$^{-3}$,
identical to our results with the old, burst model for the thick disk.  

We also explore the possibility that our model for the thin disk is in error.
Recent results by \citet{maj01} suggest that the canonical thin disk scale
height is too small. They suggest a scale height of 400-600 pc for old thin
disk populations. If their assertion is correct, the effect on our simulations
is substantial as the great majority of our WD candidates are detected at
least one scale scaleheight above the plane.   This effect would be evident
in our sample as an overabundance of thin disk white dwarfs compared to what
we expect to see with a thin disk with a scaleheight of 300 pc.

We model the \citet{maj01} thin disk as a disk with scaleheight 500 pc, $5/3$
larger then the canonical value.  To maintain consistency with the Boltzman
equations, we scale the thin disk velocity dispersions by the same factor of
$5/3$ \citep{bin98}.  This model gives an expected number of thin disk white dwarf
detections 
\begin{equation}
\nuthin = 1.9
\left( \frac{\nthin} {4.0\times10^{-3} {\rm pc}^{-3}} \right)~,
\end{equation}
and the maximum likelihood analysis gives $\nthin =
6.2\times10^{-3}$ pc$^{-3}$, $\nthick = 0.0$ pc$^{-3}$, $\nstellar = 0.0$
pc$^{-3}$, and $\ndark = 1.0\times10^{-3}$ pc$^{-3}$ with the other components
modelled as in our reference model.  Assuming a statistical error bar 
of $\sim30$\%, the \citet{maj01} thick disk scale height lowers our local thin
disk density to within $\sim1$ sigma of the canonical value.

\subsubsection{Alternative halo models}
\label{alternate_halo}

Since the stellar halo and dark halo in our models have similar density
profiles and identical kinematics, the separation of the halo signal into
the stellar halo and dark halo depends entirely on our assumptions about their
stellar content.  In \S\ref{division} we assumed a stellar halo of age 11--12
Gyr with a Saltpeter initial mass function and a dark halo of age 13--14 Gyr with
the 96IMF1 initial mass function.  This led to a result suggesting a $\sim7$\%
white dwarf dark halo.

However, this result depends entirely on our choice of initial mass function
for the dark halo.
Any assumption which makes the dark halo luminosity function any fainter will
shift the halo signal from the dark halo to the stellar halo.  For instance, a
switch to the 96IMF2 initial mass function gives a most likely answer of
$\nuthin = 3.0$, $\nuthick = 0.0$, $\nustellar=2.0$ and $\nudark=0.0$, or
$\nthin=2.4\times10^{-2}$ pc$^{-3}$,$\nthick = 0.0$ pc$^{-3}$, $\nstellar =
2.2\times10^{-4}$ pc$^{-3}$, $\ndark = 0.0$ pc$^{-3}$.  The small change in
initial mass function removes all signal from the dark halo and gives results
for the stellar halo within 1$\sigma$ of \citet{koo01}.

We note that even if our halo signal belongs in the stellar halo instead of the
dark halo, it still implies a halo white dwarf density $\sim10$ times
higher then expected from halo field star counts \citep{gou98} and still
implies that $\sim1.5$\% of the halo mass is in white dwarfs.  These estimates
are consistent with the results of \citet{opp01}.

\subsubsection{Constrained models}
\label{alternate_constraints}

In this section, we consider two constrained models in which we
fix the value of the local density for certain components and fit for the local
density of the remaining components.
Again, we do not recompute statistical error bars for
the examples in this section.  All examples in this section use the reference
models for each Galactic component described in \S~\ref{known} and~\ref{dark}.

First, we consider a solution in which we constrain the thin disk contribution
to the expected value of $\nuthin = 0.5$ (or $\nthin=4.0\times10^{-3}$
pc$^{-3}$) as given by Equation~\ref{thinscale}.  Unsurprisingly, most of the
remaining disk signal shifts into the thick disk, giving $\nthick = 1.9 \times
10^{-3}$ pc$^{-3}$, $\nstellar = 0.0$ pc$^{-3}$, $\ndark = 8.2\times10^{-4}$
pc$^{-3}$.

Next, we performed an experiment in which we adopt the \citet{koo01} approach
and first remove the thin disk contribution by hand and fit only for
contributions from the thick disk and a stellar halo.  In this approach,
we remove our brightest candidate (WD3) and fit only for the thick disk and
stellar halo. We find $\nthick = 3.2\times10^{-3}$ pc$^-3$ and $\nstellar=
6.1\times10^{-5}$ pc$^{-5}$. Allowing for a statistical error bar in our
measurements of $\sim$30\%, this is within 1.5$\sigma$ of their results.  We
also note that the \citet{koo01} point in Figure~\ref{two_component} falls
within $\sim1.5\sigma$ of our most likely point.

\subsection{Alternative WD samples}
\label{alternate_sample}

A final source of uncertainty in our analysis is the observed sample of WD
itself.  As discussed in \S\ref{selection} there is some possibility that our
WD sample should include the two marginal candidates WD6 and WD7, giving us a
7 WD sample. Taking the opposite approach we might conclude that all our faint
candidates should be excluded and include only WD1, WD3 and WD4, giving a 3 WD
sample.

\begin{figure}
\epsscale{0.5}
\plotone{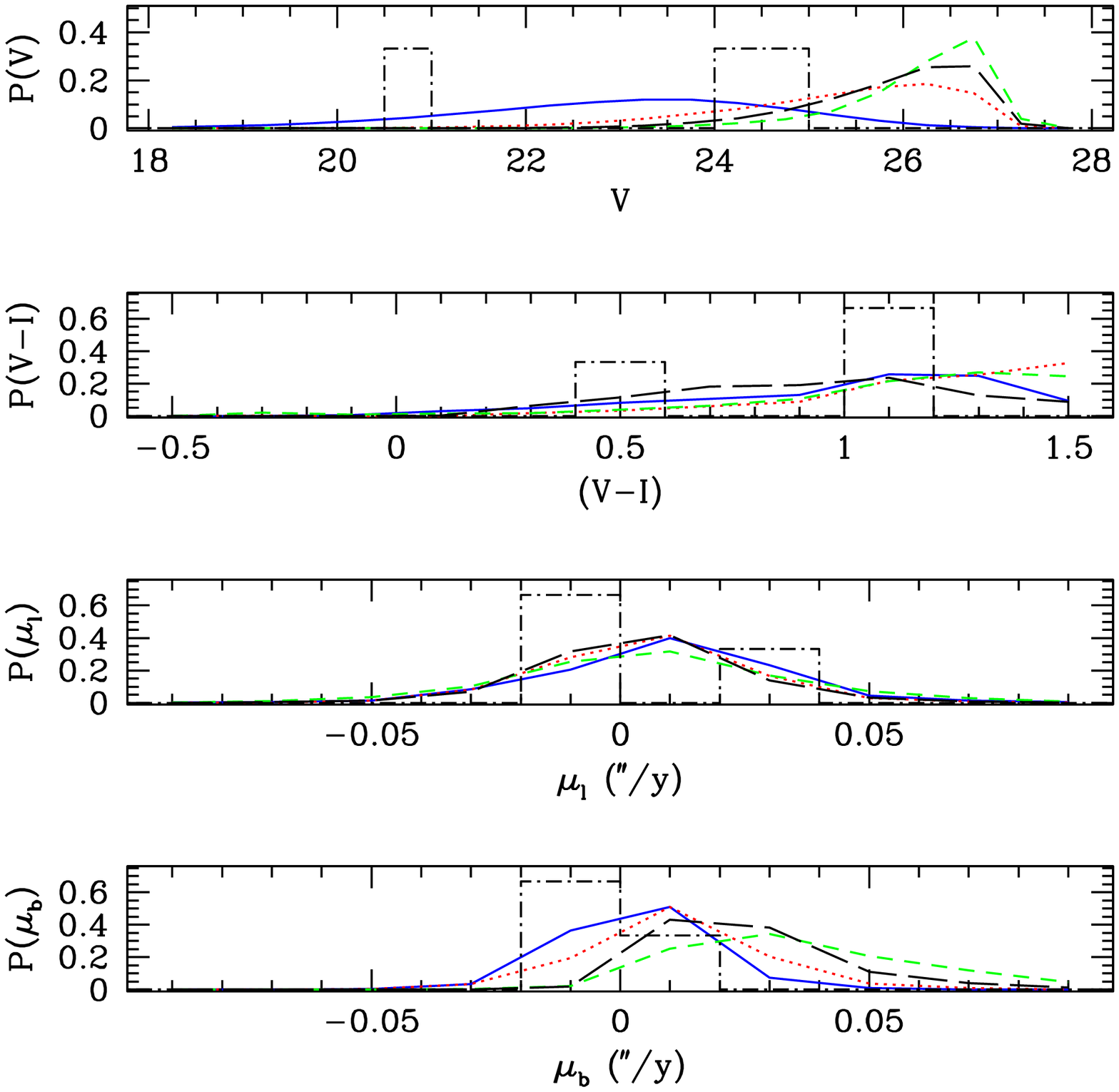}
\caption{Distributions of detectable WD for various galactic components as
described in Figure~\ref{wd_distrib}.  We overplot the distribution of 
observed WD in the 3 WD sample as a dashed black histogram.
}
\label{3wd}
\end{figure}

In Figure~\ref{3wd} we compare the 3 WD sample to the one dimensional
distribution functions of our reference model using the 96IMF1 dark halo
luminosity function.  This sample gives $(\nuthin$, $\nuthick,$ 
$\nustellar$, $\nudark)$ $= (3.0,0.0,0.0,0.0)$ for a thin disk density  $\nthin
= 2.4^{+0.7}_{-0.6}\times10^{-2}$ pc$^{-3}$.

\begin{figure}
\epsscale{0.5}
\plotone{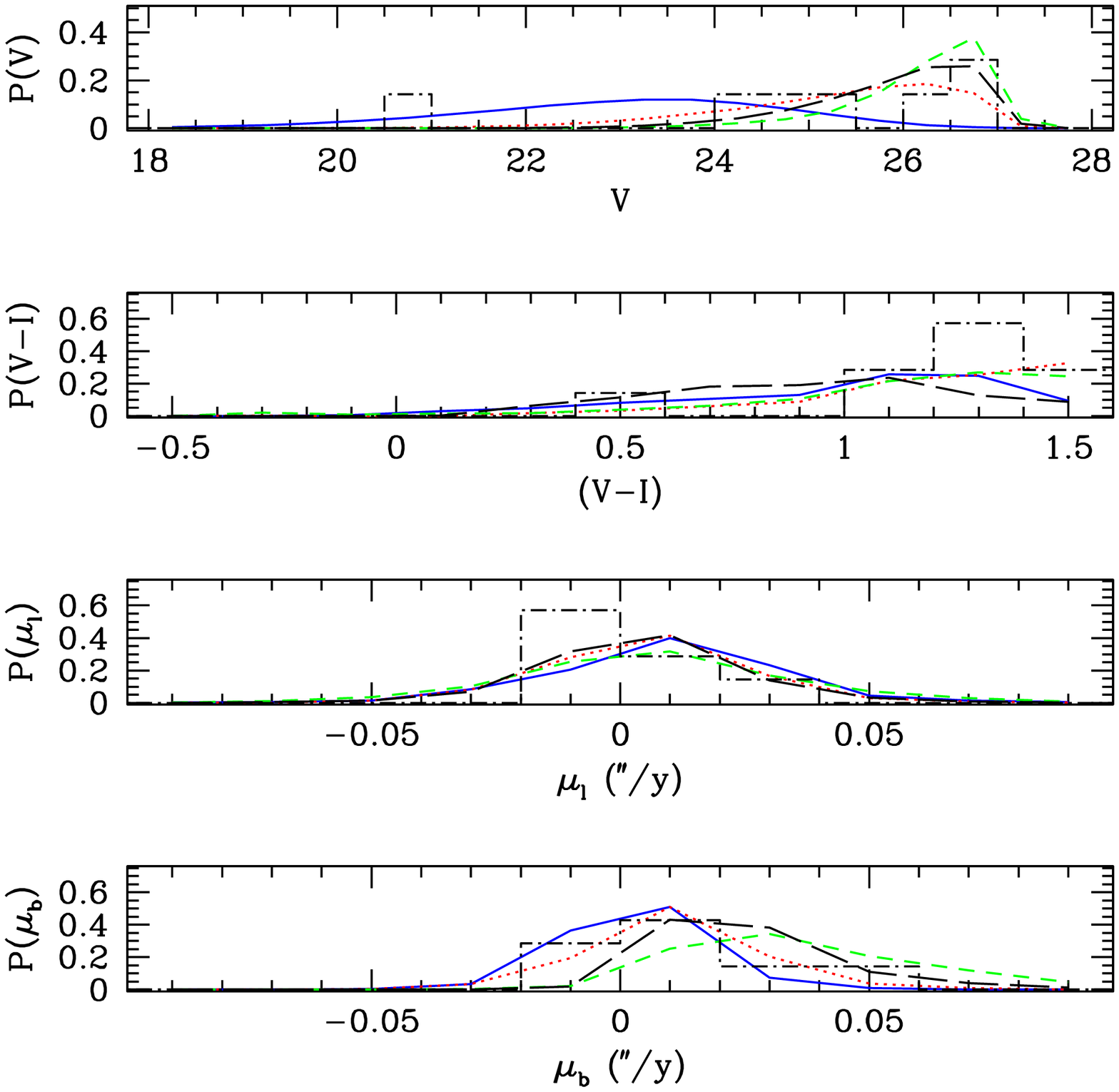}
\caption{Distributions of detectable WD for various galactic components as
described in Figure~\ref{wd_distrib}.  We overplot the distribution of 
observed WD in the 7 WD sample as a dashed black histogram.
}
\label{7wd}
\end{figure}

In Figure~\ref{7wd} we compare the 7 WD sample to the one dimensional 
distribution functions.  This sample gives $(\nuthin,\nuthick,\nustellar,\nudark) 
= (2.0,3.5,0.0,1.7)$.  The thin disk contribution is lower for the larger
sample because the outlier at $V=20.7$ is less significant.  That is, in
one dimension, this outlier accounts for only 14\% of the area under the
observed WD $V$ magnitude histogram instead of 20\% or 33\%. The maximum likelihood
analysis can put less ``power'' in the thin disk and still fit this 
outlier.  The extra power in this analysis falls entirely into the
thick disk. The 7 WD sample gives densities of $\nthin = 1.5\times10^{-2}$
pc$^{-3}$, $\nthick = 3.3\times10^{-3}$ pc$^{-3}$, $\nstellar = 0.0$ pc$^{-3}$,
$\ndark = 9.6\times10^{-5}$ pc$^{-3}$, or a 6\% WD halo.

\section{Discussion}

\citet{koo01} and \citet{rei01} demonstrate that the \citet{opp01} results do
not necessarily imply the presence of a substantial white dwarf dark matter
halo.  They also note that due to the color turnoff of ancient white dwarfs a
survey with such a  bright limiting magnitude cannot exclude the
presence of an ancient white dwarf halo of age $\gg10$ Gyr.  To exclude an
ancient halo requires a survey with a substantially fainter limiting
magnitude, such as this one.  

For faint or ancient white dwarfs, spectroscopic velocities are
not available, parallax measurements cannot be made and photometric distance
relations calibrated at bright magnitudes are potentially unreliable. Therefore,
the division of a faint white dwarf sample into disk and halo contributions
must be done without true kinematic (velocity) information.

In this work, we have explored such a statistical separation using only the
directly observable quantities of the five high proper motion white
dwarf candidates detected in a second WFPC2 epoch of the Groth-Westphal strip.
The small sample size and our imperfect knowledge of the characteristics of
the putative white dwarf dark halo lead to possible large systematic errors in
our analysis.  Using the 96IMF1 dark halo gives a 7\% white dwarf halo and
$\nthin= 2.4^{+0.5}_{-0.4}\times10^{-2}$ pc$^{-2}$, a gross excess of thin
disk white dwarfs.  However, we explore several alternative Galactic models
which demonstrate that uncertainties in our
models lead to possible systematic errors which may be larger then the
quoted statistical errors.  

For instance, the thin disk local density can be lowered four-fold and 
brought to within $1$ sigma of the canonical value by assuming the larger
thin disk scale height suggested by \citet{maj01}.  Also, if we assume
a slightly different dark halo initial mass function, the halo signal
shifts to the stellar halo where it implies a local stellar halo white 
dwarf density of $\nstellar \sim 2.2 \times 10^{-4}$, about 10 times
higher then the canonical value and similar to the results reported by
\citet{opp01}.

The use of external constraints also affects our results.
If we constrain the thin disk contribution to the canonical value, or 
attempt to remove the thin disk contribution by hand (as in \citet{koo01}),
the disk signal shifts into the thick disk, 
giving $\nthick \sim 3 \times 10^{-3}$ pc$^{-3}$.
This elevated thick disk density is similar to the results of \citet{koo01}.

However, regardless of the details of our models,
we always find $\sim3$ disk white dwarfs and $\sim2$ halo white dwarfs, a
clear excess above the $\sim$1--2 total detections expected from known
stellar populations.  We are unable to definitely determine the source of
the excess signal, but its existence seems clear.  We summarize the
local densities found in our various calculations in Table 3.

We note that the ages, abundances and kinematics of the known stellar
populations of our Galaxy are still a source of lively debate.  We have
attempted to highlight how our results depend on our assumptions about these
quantities in \S4.4.  Attempts to more precisely determine the
properties of the various components of our Galaxy will benefit from surveys
with very precise astrometry.

Both our results and those of \citet{opp01} are intriguing and together
explore the two extremes of survey philosophies: bright limiting magnitude
with many objects and faint limiting magnitude with few objects.  The
\citet{opp01} sample is bright enough to be able to employ photometric distance 
calibrations, however, fainter surveys necessary to truly exclude or confirm the
dark white dwarf halo must employ an approach more similar to ours. Future
improved knowledge of the input model characteristics and a faint sample with
more objects may reduce the model dependencies which hamper our conclusions.

\acknowledgements
We thank Andrew Dolphin for many helpful discussions on adapting HSTphot to
our specific needs and Brad Hansen for discussion of white dwarf models and
luminosity functions.  We thank Alison Vick, our STScI staff contact for
helping schedule the demanding second epoch.  Support for this publication was
provided by NASA through proposal number GO-8698, and from the Space Telescope
Science Institute, which is operated by the Association of Universities for
Research in Astronomy, under NASA contract NAS5-26555. This work was performed
under the auspices of the U.S. Department of Energy, National Nuclear Security
Administration by the University of California, Lawrence Livermore National
Laboratory under contract No. W-7405-Eng-48.  Nelson is supported in part by
an National Physical Science Consortium Graduate Fellowship.

\clearpage

\clearpage

\begin{deluxetable}{rrrr}
\tablewidth{0pt}
\tablecaption{Observations}
\tablehead{
\colhead{R.A. (2000)} & \colhead{Decl. (2000)} & \colhead{Epoch 1 (MJD)}
& \colhead{Epoch 2 (MJD)}
}
\startdata
14:15:21 & 52:02:59 & 49426.1 & 51978.7 \\ 
14:15:27 & 52:04:10 & 49419.9 & 51979.4 \\ 
14:15:33 & 52:05:20 & 49419.7 & 51991.8 \\ 
14:15:40 & 52:06:30 & 49419.4 & 51988.8 \\ 
14:15:46 & 52:07:40 & 49451.4 & 51984.0 \\ 
14:15:53 & 52:08:51 & 49426.6 & 51988.9 \\ 
14:15:59 & 52:10:01 & 49420.9 & 51978.0 \\ 
14:16:06 & 52:11:11 & 49428.8 & 51978.9 \\ 
14:16:12 & 52:12:21 & 49420.6 & 51990.1 \\ 
14:16:19 & 52:13:31 & 49422.0 & 51986.1 \\ 
14:16:25 & 52:14:41 & 49428.5 & 51978.3 \\ 
14:16:31 & 52:15:51 & 49428.0 & 51983.0 \\ 
14:16:38 & 52:17:01 & 49425.8 & 51977.1 \\ 
14:17:10 & 52:22:51 & 49418.9 & 51993.0 \\ 
14:17:17 & 52:24:01 & 49418.6 & 51994.8 \\ 
14:17:56 & 52:31:00 & 49422.8 & 51987.9 \\ 
14:18:03 & 52:32:10 & 49427.4 & 51984.2 \\ 
\enddata
\end{deluxetable}

\begin{deluxetable}{lrrrrrrrrrrr}
\rotate
\tablewidth{0pt}
\tabletypesize{\scriptsize}
\tablecaption{White Dwarf Candidates}
\tablehead{
\colhead {Candidate} & \colhead{RA} & \colhead{DEC} & \colhead{$\mu$} &
\colhead {$\mu_{l}$} & \colhead {$\mu_{b}$} &
\colhead{$V$} & \colhead {$V-I$} &
\colhead{$t_{\rm cool}$ (Gyr)} & \colhead{$M_V$} & \colhead{$d$(pc)} & \colhead{$v_{t}$(km/s)} 
} 
\startdata
WD1     &14:15:54.5&52:09:16.8& 0.020 &-0.038 &-0.008 & 24.81 & 1.09 &  $3.7^{+3.2}_{-3.1}$ & $14.9^{+0.9}_{-0.5}$ &  $940^{+370}_{-600}$ &  $90^{+20}_{-25}$ \\
        & & &       &       &       &       &      & $11.4^{+0.1}_{-0.7}$ & $17.4^{+0.2}_{-0.3}$ &  $310^{+50}_{-40}$   &  $29^{+5}_{-4}$ \\ 
WD2     &14:16:02.0&52:10:44.5& 0.029 & 0.035 & 0.024 & 26.90 & 1.38 &  $7.9^{+2.0}_{-3.3}$ & $16.1^{+0.8}_{-0.8}$ & $1460^{+800}_{-530}$ & $205^{+112}_{-74}$ \\
        & & &       &       &       &       &      & $10.5^{+0.6}_{-0.4}$ & $17.1^{+0.2}_{-0.1}$ &  $910^{+130}_{-120}$ & $128^{+18}_{-18}$ \\
WD3     &14:16:13.3&52:11:38.6& 0.032 & 0.053 & 0.018 & 20.67 & 0.55 &  $1.2^{+0.6}_{-0.5}$ & $13.4^{+0.6}_{-0.7}$ &  $290^{+100}_{-85}$  &  $44^{+24}_{-13}$ \\
        & & &       &       &       &       &      & $12.5^{+0.5}_{-0.3}$ & $17.6^{+0.1}_{-0.1}$ &   $40^{+5}_{-5}$     &   $6^{+1}_{-1}$ \\
WD4     &14:16:12.8&52:12:12.4& 0.019 &-0.038 &-0.004 & 24.02 & 1.02 &  $3.2^{+3.5}_{-1.1}$ & $14.7^{+0.8}_{-0.6}$ &  $720^{+290}_{-250}$ &  $66^{+20}_{-23}$ \\
        & & &       &       &       &       &      & $11.6^{+0.5}_{-0.6}$ & $17.4^{+0.2}_{-0.2}$ &  $210^{+40}_{-30}$   &  $19^{+3}_{-2}$ \\
WD5     &14:15:25.8&52:04:16.1& 0.051 & 0.021 & 0.050 & 25.49 & 1.28 &  $6.6^{+3.3}_{-3.1}$ & $15.7^{+1.1}_{-0.9}$ &  $900^{+500}_{-410}$ & $217^{+126}_{-91}$ \\
        & & &       &       &       &       &      & $10.8^{+0.7}_{-0.7}$ & $17.2^{+0.2}_{-0.2}$ &  $460^{+80}_{-40}$   & $111^{+20}_{-18}$ \\
WD6$^*$ &14:15:39.8&52:04:48.3& 0.016 &-0.021 & 0.012 & 26.33 & 1.62 & $10.0^{+0.3}_{-1.8}$ & $16.9^{+0.1}_{-0.7}$ & $770^{+370}_{-110}$  &  $57^{+29}_{-7}$ \\
WD7$^*$ &14:18:25.8&52:04:16.1& 0.019 &-0.038 & 0.003 & 26.93 & 1.60 & $10.0^{0.1}_{-2.0}$  & $16.9^{+0.1}_{-0.8}$ & $1000^{+560}_{-130}$ &  $91^{+49}_{14}$ \\
\enddata
\tablecomments{$^*$ indicates marginal white dwarf candidates as discussed
in \S\ref{selection}.  Proper motions in the plane of the sky, $\mu$, 
in the direction of increasing Galactic longtitude $\mu_{l}$, and latitude,
$\mu_{b}$, are given in units of $^{\prime\prime}$/yr. We estimate an
uncertainty on all $V$ magnitudes of 0.15 mag and an uncertainty on all
$(V-I)$ colors of 0.20 mag. The error in proper motion, $\mu$ is 
0.004 ${\prime\prime}$/yr. Error bars are calculated for derived quantities
by finding the maximum and minimum values attainable within 1 unit of
photometric uncertainty.}
\end{deluxetable}

\begin{deluxetable}{lrrrr}
\rotate
\tablecaption{Local white dwarf densities}
\tabletypesize{\scriptsize}
\tablehead{
\colhead{Source} & \colhead{$\nthin$} & \colhead{$\nthick$} &
\colhead{$\nstellar$} & \colhead{$\ndark$}
}
\startdata
Canonical Values                  & $4.0\times10^{-3}$ 
                                  & $9.5\times10^{-5}$ 
		                  & $2.2\times10^{-5}$ 
		                  & $0.0$ \\
\citet{koo01}                     & \nodata            
                                  & $1.8^{+0.5}_{-0.5}\times10^{-3}$ 
		                  & $1.1^{2.1}_{-0.7}\times10^{-4}$ 
		                  & \nodata \\    
Reference Model (\S 4.3)          & $2.4^{+0.7}_{-0.6}\times10^{-3}$ 
                                  & $0.0^{+7.6}\times10^{-4}$ 
				  & $0.0^{+7.7}\times10^{-5}$ 
				  & $1.0^{+0.4}_{-0.4}\times10^{-3}$ \\ 
Increased Thin Disk Scale Height(\S 4.4.1) & $6.2\times10^{-3}$
                                  & $0.0$
				  & $0.0$
				  & $1.0\times10^{-3}$\\
96IMF2 Dark Halo (\S 4.4.2)       & $2.4\times10^{-3}$
                                  & $0.0$
				  & $2.2\times10^{-4}$
				  & $0.0$\\
Contrained Thin Disk (\S 4.4.3)   & $4.0\times10^{-3}$
                                  & $1.9\times10^{-3}$
				  & $0.0$
				  & $8.2\times10^{-4}$\\
\enddata
\tablecomments{All densities are given in units of $M_\odot$ pc$^{-3}$.  The
known stellar populations assume a mean white dwarf mass of 0.6 $M_\odot$. The
dark halo assumes a mean white dwarf mass of 0.5 $M_\odot$. The \citet{koo01}
work does not distinguish between a stellar and a dark halo.  We have listed
their halo density under the $\nstellar$ column, but it could also be listed
under $\ndark$.}
\end{deluxetable}
\end{document}